\documentclass[preprint,nofootinbib,aps,superscriptaddress,eqsecnum]{revtex4-1}
\pdfoutput=1
\usepackage[colorlinks=true,citecolor=blue,linkcolor=purple,breaklinks=true]{hyperref}
\usepackage{graphicx}
\usepackage{mathrsfs}
\usepackage{booktabs}
\usepackage{enumerate}
\usepackage{slashed}
\usepackage{latexsym, amssymb} 
\usepackage{amsmath,subfigure,xcolor}
\usepackage{soul}

\newcommand{\cw}{\rm{CW}}
\newcommand{\ordone}{\mathcal{O}(1)}
\newcommand{\vev}{{\textit{vev}} }
\newcommand{\vevs}{{\textit{vev}}s }
\newcommand{\tr}{{\rm{Tr}}}

\newcommand{\beq}{\begin{equation}}
\newcommand{\eeq}{\end{equation}}
\newcommand{\bea}{\begin{eqnarray}}
\newcommand{\eea}{\end{eqnarray}}

\newcommand{\nn}{\nonumber}

\newcommand{\gbl}{g_{BL}}
\def\a{\alpha}

\def\k{\kappa}

\def\l{\lambda}
\def\m{\mu}
\def\n{\nu}

\begin{document}
\title{Gravitational Wave imprints of the \\Doublet Left-Right Symmetric Model}

\author{Siddhartha Karmakar\footnote{karmakars@iitb.ac.in}}
\affiliation{{\small Department of Mathematics, Indian Institute of Technology Bombay,\\
 Powai, Mumbai, Maharashtra-400076, India}}
\affiliation{{\small Department of Theoretical Physics, Tata Institute of Fundamental Research,\\ Colaba, Mumbai, Maharashtra-400005, India}}
\author{Dhruv Ringe\footnote{phd1901151004@iiti.ac.in}}
\affiliation{{\small Department of Physics, Indian Institute of Technology Indore, \\Khandwa Road, Simrol, Indore, Madhya Pradesh-453552, India}}

\begin{abstract}
{\noindent
We study the stochastic gravitational wave~(GW) background resulting from the strong first-order phase transition~(SFOPT) associated with $SU(2)_R\times U(1)_{B-L}$-breaking in the doublet left-right symmetric model~(DLRSM). For different values of the symmetry-breaking scale $v_R =20,~30$, and $50$ TeV, we construct the one-loop finite temperature effective potential to explore the parameter space for regions showing SFOPT. We identify the region where the associated GW background is strong enough to be detected at planned GW observatories. A strong GW signature favors a relatively light CP-even neutral scalar $H_{3}$, arising from the $SU(2)_R$ doublet. The $SU(2)_L$ subgroup of DLRSM is broken by three {\it vevs}: $\k_1,~\k_2$, and $v_L$. We observe a preference for $\mathcal{O}(1)$ values of the ratio $w=v_L/\k_1$, but no clear preference for the ratio $r=\k_2/\k_1$. A large number of points with strong GW signal can be ruled out from precise measurement of the trilinear Higgs coupling and searches for $H_3$ at the future colliders. }
\end{abstract}

\pacs {}
\maketitle

\section{Introduction}
Left-right symmetric models\,(LRSMs)\,\cite{Pati:1974yy,Mohapatra:1974gc,PhysRevD.11.566,Senjanovic:1975rk,Senjanovic:1978ev} provide an attractive scenario for addressing several limitations of the standard model (SM). In LRSM, the SM gauge group is extended from $\mathcal{G}_{\rm{SM}} = SU(3)_c\times SU(2)_L\times U(1)_{Y}$ to $\mathcal{G}_{\rm{LRSM}} = SU(3)_c\times SU(2)_L\times SU(2)_R\times U(1)_{B-L}$, and the right-chiral fermions transform as doublets under the $SU(2)_R$ subgroup. The non-observation of a right-handed charged current at colliders puts a lower bound on the $SU(2)_R\times U(1)_{B-L}$-breaking scale, $v_R\gtrsim \mathcal{O}(10)$\,TeV, while the upper bound remains unconstrained. 

There are different realizations of LRSM, depending on the scalars involved in the spontaneous breaking of $\mathcal{G}_{\textrm{LRSM}}$ to $\mathcal{G}_{\rm{SM}}$. These also differ in the mechanism for generating fermion masses. The triplet left-right symmetric model\,(TLRSM)\,\cite{Maiezza:2016ybz,PhysRevD.44.837,Senjanovic:2016bya}, contains a scalar bi-doublet, and two $SU(2)$ triplets. The charged fermion masses are then generated by the bi-doublet, whereas the neutrino masses are generated by the type-II seesaw mechanism\,\cite{Schechter:1981cv}. On the other hand, the scalar sector of a doublet left-right symmetric model\,(DLRSM) consists of a scalar bi-doublet and a pair of $SU(2)$ doublets\,\cite{Senjanovic:1978ev,Mohapatra:1977be}. In DLRSM, neutrino masses can be incorporated by extending the model with an additional charged singlet scalar\,\cite{Babu:1988qv, FileviezPerez:2016erl, FileviezPerez:2017zwm, Babu:2020bgz}. 
Contrary to TLRSM, where the \vev of the triplets is constrained to be small, there are no sources of custodial symmetry breaking in DLRSM at the tree level. Apart from TLRSM and DLRSM, other variations have also been discussed in the literature and have different experimental consequences\,\cite{Ma:2012gb,Frank:2020odd,Graf:2021xku,Akhmedov:1995vm}. 

The era of gravitational wave\,(GW) astronomy was kickstarted by the observation of GW from a binary black hole merger by the aLIGO collaboration in 2015\,\cite{Abbott_2016}. Several ground-based and space-based observatories such as LISA\,\cite{LISA:2017pwj}, DECIGO\,\cite{Seto:2001qf}, BBO\,\cite{Corbin:2005ny}, ET\,\cite{Punturo:2010zz}, and CE\,\cite{LIGOScientific:2016wof} are planned and will be functional in the coming decades. Various phenomena in the early universe, such as inflation, cosmic strings, domain walls, and strong first-order phase transitions\,(SFOPT) can lead to a stochastic GW background\,\cite{Caprini:2015zlo,Athron:2023xlk,LISACosmologyWorkingGroup:2022jok,Caprini:2018mtu}. The
upcoming GW observatories will be capable of detecting the GW background from SFOPT upto symmetry-breaking scales as high as $10^6- 10^7$\,GeV\,\cite{Dev:2016feu,Addazi:2018nzm,Ringe:2022rjx}. 

In the context of LRSM, GW astronomy presents a novel approach to probe the scale $v_R$ by studying the possibility of an observable GW background from SFOPT within the LRSM.  Different realizations of LRSM have been explored in the literature for GW imprints: through SFOPT in TRLSM\,\cite{Brdar:2019fur, Li:2020eun}, and in an LR model with seesaw-like fermion mass generation\,\cite{Graf:2021xku}, and also from domain walls arising out of the breaking of the discrete parity symmetry\,\cite{Borah:2022wdy}. However, GW imprints of DLRSM have not yet been explored in the literature. Recently it was shown that the pattern of electroweak symmetry breaking (EWSB) in DLRSM can be vastly  different from the other versions of LRSM, with interesting consequences from precision observables\,\cite{Bernard:2020cyi} and Higgs data\,\cite{Karmakar:2022iip}. It is therefore interesting to study the GW signature in DLRSM.

EWSB in DLRSM happens via three \vevs: $\k_1,~\k_2$, coming from the bi-doublet, and $v_L$ coming from the $SU(2)_L$ doublet. These are constrained by the relation, $\k_1^2+\k_2^2+v_L^2=v^2$, where $v = 246$\,GeV. It is useful to define the \vev ratios: $r = \k_2/\k_1,~w = v_L/\k_1$. As the custodial symmetry is preserved at the tree level, the \vevs $\k_1,~ \k_2$, and $v_L$ can all be sizable, i.e. $r$ and $w$ can be $\mathcal{O}(0.1)$ and $\ordone$ respectively. In ref.\,\cite{Bernard:2020cyi}, it was shown that the EW precision data prefers a large value of $w$. Further, in ref.\,\cite{Karmakar:2022iip} it was shown that the measurement of the Higgs signal strength and meson mixing bounds prefer large values of $r$ and $w$. It is therefore interesting to note that, unlike TLRSM, EWSB in DLRSM can be considerably different from that in SM, even though the $SU(2)_R\times U(1)_{B-L}$-breaking dynamics is decoupled from the EW scale. In this paper, we ask \textbf{(i)} whether DLRSM can lead to a detectable GW background in some region of the parameter space, and \textbf{(ii)} whether this region of the parameter space prefers a special pattern of EWSB.

The rest of the paper is organised as follows. In Sec.\,\ref{Sec:Model}, we give a brief review of DLRSM: field content, symmetry breaking, and mass generation in the gauge, fermion, and gauge sectors. In Sec.\,\ref{subsec: theory} and Sec.\,\ref{subsec: higgs}, we discuss the theoretical bounds and the constraints from the Higgs data respectively. In Sec.\,\ref{sec: effective potential}, we construct the one-loop finite temperature effective potential required to study the phase transition (PT) associated with $SU(2)_R\times U(1)_{B-L}$-breaking. We then describe our procedure for scanning the parameter space in Sec.\,\ref{sec: paramscan}. In Sec.\,\ref{sec: GW}, we discuss the GW background obtained for points with SFOPT. In Sec.\,\ref{sec: detection prospects}, we compute the signal-to-noise ratio (SNR) for six benchmark points, at various planned GW detectors such as FP-DECIGO, BBO, and Ultimate-DECIGO. In Sec.\,\ref{Sec: collider}, we discuss future collider probes that can complement the GW signal. Finally, in Sec.\,\ref{sec: summary}, we summarize our key findings and present concluding remarks.

\section{The model}
\label{Sec:Model}
We follow the notation of refs.\,\cite{Bernard:2020cyi,Karmakar:2022iip} for the scalar potential and \vev structure of the scalar multiplets. The fermion content of the model has the following charges under the LRSM gauge group, $\mathcal{G}_{\rm{LRSM}} = SU(3)_c \times SU(2)_L\times SU(2)_R \times U(1)_{B-L}$,
\bea
Q_L &=& \begin{pmatrix}
 u_L \\
 d_L 
\end{pmatrix} \sim (3,2,1,1/3), \hspace{10mm}
Q_R = \begin{pmatrix}
 u_R \\ 
 d_R 
\end{pmatrix} \sim (3,1,2,1/3), \nn \\
L_L &=& \begin{pmatrix}
 \nu_L \\
 e_L 
\end{pmatrix} \sim (1,2,1,-1), \hspace{12mm}
L_R = \begin{pmatrix}
 \nu_R \\
 e_R 
\end{pmatrix} \sim (1,1,2,-1),
\label{eq:fermion}
\eea
where the quantum numbers of the  multiplets under the sub-groups of $\mathcal{G}_{\rm{LRSM}}$ are indicated in brackets. We have suppressed the family index $i\in\{1,2,3\}$ for three generations of quarks and three generations of leptons.The right-handed neutrino $\nu_R$ is needed to complete the $SU(2)_R$ lepton doublet. This choice of fermions is required for the cancellation of the $U(1)_{B-L}$ gauge anomaly and ensures that the model is manifestly symmetric under the transformations: $Q_L \leftrightarrow Q_R$, $L_L \leftrightarrow L_R$. 

\subsection{Scalar sector}
\label{sec:ScalarSector}
The scalar sector of DLRSM includes a complex bi-doublet $\Phi$ needed to generate charged fermion masses, and two doublets $\chi_L$ and $\chi_R$, which participate in the EW- and LR-symmetry breaking respectively. These scalar multiplets and their charges under $\mathcal{G}_\text{LRSM}$ are:
\bea
\Phi = \begin{pmatrix}
    \phi_1^0 & \phi_2^+ \\
    \phi_1^- & \phi_2^0
\end{pmatrix} \sim (1,2,2,0), ~ \chi_L = \begin{pmatrix}
    \chi_L^+ \\
    \chi_L^0
\end{pmatrix} \sim (1,2,1,1), ~ \text{and}~\chi_R = \begin{pmatrix}
    \chi_R^+ \\
    \chi_R^0
\end{pmatrix} \sim (1,1,2,1).\nn\\
\eea

We take the potential to be parity-symmetric, i.e. the couplings of `L' and `R' fields are equal. This imposes an additional discrete symmetry $\mathcal{P}: L\leftrightarrow R$ on the Lagrangian. The most general, CP-conserving, renormalizable scalar potential is then given by,
\bea\label{eq: potential}
V  &=& V_2 + V_3 + V_4,\nn\\
V_2  &=& -\m_1^2\tr(\Phi^{\dagger}\Phi) - \m_2^2\ [\tr(\tilde{\Phi}\Phi^{\dagger})+ \tr(\tilde{\Phi}^{\dagger} \Phi)] - \m_3^2\ [\chi_L^{\dagger} \chi_L + \chi_R^{\dagger} \chi_R], \nn\\
V_3 &=& \m_4\  [\chi_L^{\dagger} \Phi \chi_R + \chi_R^{\dagger} \Phi^{\dagger} \chi_L] + \m_5\  [\chi_L^{\dagger} \tilde{\Phi} \chi_R + \chi_R^{\dagger}\tilde{\Phi}^{\dagger}\chi_L ], \nn\\
V_4 &=& \l_1[\tr(\Phi^{\dagger}\Phi)]^2 + \l_2\ [ [\tr(\tilde{\Phi} \Phi^{\dagger})]^2
 + [\tr(\tilde{\Phi}^{\dagger} \Phi)]^2 ]  + \l_3\text{Tr}(\tilde{\Phi} \Phi^{\dagger}) \, \tr(\tilde{\Phi}^{\dagger} \Phi)  \nn\\
 &&+ \l_4\tr(\Phi^{\dagger}\Phi) \, [\tr(\tilde{\Phi}\Phi^{\dagger})+ \tr(\tilde{\Phi}^{\dagger}\Phi)]  + \rho_1\  [(\chi_L^{\dagger} \chi_L )^2 + (\chi_R^{\dagger} \chi_R )^2]
 + \rho_2\  \chi_L^{\dagger} \chi_L \chi_R^{\dagger}\chi_R \nn\\
 &&+ \alpha_1\tr(\Phi^{\dagger} \Phi ) [\chi_L^{\dagger}\chi_L + \chi_R^{\dagger}\chi_R ]
 + \Big\{\frac{\alpha_2}{2} \ [\chi_L^{\dagger} \chi_L  \tr(\tilde{\Phi} \Phi^{\dagger} ) + \chi_R^{\dagger} \chi_R  \tr(\tilde{\Phi}^{\dagger} \Phi )] + {\rm h.c.} \Big\} \nn\\
 &&+ \alpha_3\ [\chi_L^{\dagger}\
 \Phi \Phi^{\dagger}\chi_L + \chi_R^{\dagger} \Phi^{\dagger} \Phi  \chi_R  ] 
 + \alpha_4\ [\chi_L^{\dagger}\
 \tilde{\Phi} \tilde{\Phi}^{\dagger}\chi_L + \chi_R^{\dagger} \tilde{\Phi}^{\dagger} \tilde{\Phi}  \chi_R  ],
 \label{eq:scalarpotential}
\eea
with $\tilde{\Phi}\equiv \sigma_2\Phi^*\sigma_2$. The potential has mass parameters: $\{\m_{1,2,3,4,5}\}$, and quartic couplings: $\{\l_{1,2,3,4},\a_{1,2,3,4},\rho_{1,2}\}$. We assume all parameters to be real for simplicity. 

The neutral scalars can be written in terms of real and imaginary components,
\bea
\phi_1^0 &=& \frac{1}{\sqrt{2}} (\phi_{1r}^0 + i \phi_{1i}^0), \,\,\,\,
\phi_2^0 = \frac{1}{\sqrt{2}} (\phi_{2r}^0 + i \phi_{2i}^0), \nn\\
\chi_L^0 &=& \frac{1}{\sqrt{2}} (\chi_{Lr}^0 + i \chi_{Li}^0), \,\,\,\,
\chi_R^0 = \frac{1}{\sqrt{2}} (\chi_{Rr}^0 + i \chi_{Ri}^0). 
\eea
We assign non-zero \vevs only to the real components of the neutral scalars and do not consider CP- or charge-breaking minima. The \vev structure is denoted by
\beq
     \langle\Phi \rangle = \frac{1}{\sqrt{2}}\begin{pmatrix}
     \k_1 & 0\\
     0 & \k_2
 \end{pmatrix}, ~ \langle\chi_L\rangle = \frac{1}{\sqrt{2}} \begin{pmatrix}
     0\\
     v_L
 \end{pmatrix}, ~ \langle\chi_R \rangle = \frac{1}{\sqrt{2}}\begin{pmatrix}
     0\\
     v_R
 \end{pmatrix}\,.
 \eeq

The pattern of symmetry breaking is as follows:
$$SU(2)_L\times SU(2)_R \times U(1)_{B-L}\xrightarrow{\mathit{~~v_R~~}} SU(2)_L\times U(1)_{Y} \xrightarrow{\mathit{\k_1,\k_2,v_L}} U(1)_Y . $$
The \vev $v_R$ of the doublet $\chi_R$ breaks $SU(2)_R\times U(1)_{B-L}$, while the three \vevs $\k_1,~\k_2$, and $v_L$ trigger EWSB. Note that the discrete LR symmetry $\mathcal{P}$, is also broken by $v_R$, which leads to the formation of domain walls \cite{Borah:2022wdy,Chakrabortty:2019fov,Mishra:2009mk,Borah:2011qq,Banerjee:2020zxw,Borboruah:2022eex}. Such a domain wall network can dominate the energy density of the universe at late times. To avoid domination, a small bias term can be introduced via explicit LR-breaking operators, so that the domain walls become unstable and decay before the epoch of big bang nucleosynthesis. For example, the bias was generated by Planck-suppressed higher dimensional operators in ref.\,\cite{Borah:2022wdy}. Due to large suppression, these do not affect the nature and strength of the $SU(2)_R \times U(1)_{B-L}$ breaking PT.

As mentioned earlier, the EW \vevs can be conveniently expressed in terms of the \vev ratios $r$ and $w$ as, $\k_2 = r \k_1$ and $v_L = w \k_1$.
Then, $\k_1^2 (1 + r^2 + w^2) = v^2$, i.e., the value of $\k_1$ is fixed for a given $r$ and $w$. The absence of a right-handed charged current in collider experiments implies a hierarchy of scales $v_R\gg v$. 

In terms of the \vevs $\k_1, \k_2, v_L$, and $v_R$, the minimization conditions are,
\beq
\frac{\partial V}{\partial \k_1} = \frac{\partial V}{\partial \k_2} = \frac{\partial V}{\partial v_L} = \frac{\partial V}{\partial v_R} = 0.
\eeq
Using the minimization conditions, we trade $\m_1^2$, $\m_2^2$, $\m_3^2$, and $\m_5$ for the \vevs, $\m_4$, and quartic couplings (see Appendix \ref{appendix: min} for full expressions). Thus the parameters of the DLRSM scalar sector reduce to
\bea 
\{\l_{1,2,3,4}, \a_{1,2,3,4}, \rho_{1,2}, \m_4, r, w, v_R \}\,\, .
\label{eq: scalar_sector_params}
\eea

The CP-even, CP-odd, and charged scalar mass matrices are obtained using 
\beq
m^2_{ij} = \left.\frac{\partial^2 V}{\partial \varphi_i \,\partial\varphi_j}\right\vert_{\langle \varphi \rangle},
\eeq
where
\beq\label{eq: varphi}
\varphi \equiv \{\phi^0_{1r},\phi^0_{2r},\chi^0_{Lr},\chi^0_{Rr},\phi^0_{1i},\phi^0_{2i},\chi^0_{Li},\chi^0_{Ri},\phi^{\pm}_{1},\phi^{\pm}_{2},\chi^{\pm}_{L},\chi^{\pm}_{R}\}\,\,.
\eeq
Physical scalar masses and mixing angles are obtained by diagonalizing these matrices. We denote the physical spectrum of scalars by:
 CP-even scalars, $h,~ H_1,~ H_2,~ H_3$,
 CP-odd scalars, $A_1,~A_2$, and the 
charged scalars, $H_1^{\pm},~H_2^{\pm}$. 

The lightest CP-even scalar, $h$ has a mass of the order $v$, and is identified with the SM-like Higgs with mass $\sim 125$ GeV. Using non-degenerate perturbation theory, $m_h$ is estimated as\,\cite{Bernard:2020cyi,Karmakar:2022iip}
\bea 
m_{h, \text{analytic}}^2 &=& \frac{\k_1^2}{2 (1+r^2 +w^2)}\times\nn\\
&&\Bigg( 4 \Big(\l_1 (r^2+1)^2 + 4 r (\l_4(r^2+1)+r \l_{23}) + w^2 (\a_{124} + r^2 (\a_1+\a_3) \nn\\ &&+\a_2 r) + \rho_1 w^4 \Big) 
 -\frac{1}{\rho_1}(\a_{124} + r^2 (\a_1+\a_3) + \a_2 r + 2 \rho_1 w^2)^2 \Bigg)\,,
\label{eq:mh_analytic}
\eea
where, $\a_{124} \equiv \a_1+r\a_2+\a_4$, and $\l_{23} = 2\l_2+\l_3$. In the limit $r,w\rightarrow 0$, the above expression simplifies to
\beq\label{eq: mH_approx}
m_{h, \text{analytic}}^2 = v^2\left(2\l_1 - \frac{(\a_1+\a_4)^2}{2\rho_1}\right).
\eeq
However, it was pointed out in ref.\,\cite{Karmakar:2022iip} that for certain values of the quartic parameters, the analytical estimate for $m_h$ may not suffice. 

The other scalars have masses of the order $v_R$. To $\mathcal{O}\big(\k_1/v_R\big)$, these masses are related to each other as
\bea
&m_{H_1}^2\simeq m_{A_1}^2\simeq m_{H^{\pm}_1}^2 \approx \frac{1}{2}(\a_3-\a_4) v_R^2\,, &\nn\\
&m_{H_2}^2\simeq m_{A_2}^2\simeq m_{H^{\pm}_2}^2 \approx \frac{1}{2}(\rho_2 - 2 \rho_1)v_R^2\,, &\nn \\
&m_{H_3}^2 = 2\rho_1 v_R^2\,, &\nn \\
&m_{H_2}^2>m_{H_1}^2\,.&\nn 
\eea
The first two mass expressions are valid in the limit $r,w\rightarrow 0$. Positive-definite nature of the CP-even mass matrix leads to two approximate criteria: $\rho_2 > 2 \rho_1$ and $\a_3 > \a_4$. In our analysis, we calculate the scalar masses and mixing numerically. The full analytic expressions at the leading order can be found in the Appendix of ref.\,\cite{Karmakar:2022iip}. 

For the CP-even scalars, the mass-squared matrix is diagonalized by the orthogonal matrix $O$, 
\bea 
O^T \mathcal{M}^2_\text{CPE} O = \big(\mathcal{M}^{\text{diag}}_\text{CPE}\big)^2,\,\,\,\,\,\, X_\text{physical} = O^T X\, ,
\label{eq:ortho}
\eea
where $X = (\phi_{1r}^0, \phi^0_{2r}, \chi^0_{Lr}, \chi_{Rr}^0)^T$, $X_\text{physical} = (h, H_1, H_2, H_3)^T$. The scalars $H_1$ and $A_1$ can contribute to the mixing of $K_0 - \bar{K}_0$ system, leading to the constraint, $m_{H_1, A_1} > 15$\,TeV \,\cite{Zhang:2007da}. The scalar $H_3$ predominantly originates from the doublet $\chi_R$ and its coupling to the SM particles is suppressed by the element $O_{41} \sim v^2/v_R^2$. So, collider searches allow it to be much lighter than $H_1$. 

The triple Higgs coupling $(c_{h^3})$ in DLRSM is given by\,\cite{Karmakar:2022iip}
\bea\label{eq: chhh} 
 c_{h^3} &=& \frac{\k_1}{2} \Big(2 (\l_1 + r \l_4) O_{11}^3 + 2(r\l_1 + \l_4) O_{21}^3 + 2 w \rho_1 O_{31}^3 + 2(r(\l_1+4\l_2 + 2\l_3)+3\l_4) O_{11}^2 O_{21} \nn \\
 && \hspace{20pt} 
 +  2(\l_1+4\l_2 + 2\l_3 +3 \l_4 r) O_{11} O_{21}^2 + w (\a_1 + \a_4) O_{11}^2 O_{31}  + (\a_1 + r \a_2 + \a_4) O_{11} O_{31}^2 \nn\\
 &&  \hspace{20pt} 
 + w (\a_1 + \a_3) O_{21}^2 O_{31} + (\a_2 + r(\a_1 + \a_3)) O_{21} O_{31}^2   \Big) \,\,\, ,
 \label{kappah}
\eea
with the corresponding coupling multiplier $\kappa_h = c_{h^3}/c_{h^3}^\textrm{SM}$, where $c_{h^3}^\textrm{SM} = m_h^2/2 v = \l^{\rm{SM}} v$.

\subsection{Fermion sector}
The fermion multiplets couple to the bi-doublet $\Phi$ via Yukawa terms:
\bea 
\mathcal{L}_{\rm{Y}} \supset - \bar{Q}_{Li} (y_{ij} \Phi + \tilde{y}_{ij} \tilde{\Phi}) Q_{Rj} + \rm{h.c.} \,\, , 
\label{eq:uvyukawaterms}
\eea
which leads to the mass matrices for the quarks:
\bea 
M_U = \frac{1}{\sqrt{2}}(\k_1 y + \k_2 \tilde{y}), \,\,\,\,
M_D = \frac{1}{\sqrt{2}}(\k_2 y + \k_1 \tilde{y})\,\,,   \nn
\eea 
where $M_U$ and $M_D$ stand for up-type and down-type mass matrices in the flavor basis respectively. To obtain the physical basis of fermions, these mass matrices need to be diagonalized through unitary transformations described by the left- and right-handed CKM matrices ($V_{L,R}^\textrm{CKM}$). Manifest left-right symmetry implies $V_{R}^\textrm{CKM} = V_{L}^\textrm{CKM}$.
For the calculation of the effective potential in the next section, it is enough to take $y\approx\text{diag}(0,0,y_{33})$ and $\tilde{y}\approx\text{diag}(0,0,\tilde{y}_{33})$. In the limit $V^\textrm{CKM}_{33}\approx 1$, 
\bea\label{eq: y_33}
y_{33} &=& \frac{\sqrt{2}(1+r^2+w^2)^{1/2}}{v(1-r^2)}(m_t-r m_b), \nn\\
\tilde{y}_{33} &=& \frac{\sqrt{2}(1+r^2+w^2)^{1/2}}{v(1-r^2)}(m_b-r m_t),
\eea
where the top and bottom quark masses are  $m_t = 173.5$ GeV, and $m_b \approx 5$ GeV. In the limit $r,w\rightarrow 0 $, $y_{33}$ and $\tilde{y}_{33}$ reduce to the SM Yukawa couplings $y_t$ and $y_b$ respectively. However, we do not make any such assumption and use eq.\,\eqref{eq: y_33}, allowing $r,w$ to be arbitrary. 

The couplings of the SM-like Higgs with the third-generation quarks are given by: 
\bea \label{eq: hff}
c_{htt\,(hbb)} = \frac{\k_1}{\k_{-}^2}\,\Big((O_{11}-r O_{21})m_{t(b)} + (O_{21}-r O_{11}) (V_{L}^{\text{CKM}}\,\hat{M}_{D(U)}\,V_{R}^{\text{CKM} \dagger})_{33} \Big), 
\label{eq:htthbb}
\eea 
where $\k^2_- = \k_1^2-\k_2^2 = \k_1^2(1-r^2)$ and $\hat{M}_{U(D)}$ denotes the diagonal up\,(down)-type quark mass matrix. Here $O_{ij}$ are the elements of the orthogonal transformation matrix appearing in eq.\,\eqref{eq:ortho}. Then the coupling multipliers, $\k_b$ and $\k_t$ are: $\k_f = c_{hff}/c^{\rm{SM}}_{hff}$, where $c_{hff}^{\rm{SM}} = m_f/v$ and $f = t, b$.

Since $V_{L,R}^{\rm{CKM}}\approx \mathbf{1}$, eq.\,\eqref{eq: hff}, becomes,
\bea
c_{htt} &\approx & \frac{\k_1}{\k^2_-}\big(O_{11}(m_t-r m_b) + O_{21}(m_b-r m_t)\big),\nn\\
c_{hbb} &\approx & \frac{\k_1}{\k^2_-}\big(O_{11}(m_b - r m_t) + O_{21}(m_t - r m_b)\big)\nn
\eea
Note that there is a hierarchy, $O_{21}\ll O_{11}\sim 1$, $m_t\gg m_b$, and $r\ll 1$. The SM couplings are recovered by setting $O_{11}=1, O_{21}=0, r=0$, in the above expressions. For a large $\phi_{1r}^0- \phi_{2r}^0$ mixing, i.e. $O_{21} \gtrsim \mathcal{O}(10^{-2})$ or large $\k_2$, i.e. $r \sim \mathcal{O}(10^{-1})$, the deviation of $hb\bar{b}$ coupling from the SM value can be quite large due to the multiplicative factors proportional to $O_{21}m_t$, and $rO_{11}m_t$. On the other hand, the deviation of $ht\bar{t}$ coupling is proportional to $O_{21}m_b$ and $r O_{21} m_t$, and is therefore rather small for the current precision of $\k_t$ measurement.

The Yukawa term for leptons is similar to that of  quarks given in eq. (\ref{eq:uvyukawaterms}). However, since neutrino masses are tiny, generating them in DLRSM would lead to a large hierarchy among lepton Yukawa couplings. Moreover, the neutrinos could be Majorana, in which case DLRSM cannot account for them. In refs.  \cite{FileviezPerez:2016erl, FileviezPerez:2017zwm, Babu:1988qv, Babu:2020bgz}, neutrino masses were explained by adding a singlet charged scalar to DLRSM. In Appendix\,\ref{sec: neutrino mass}, we show that this extra field does not modify the strength of FOPT.

\subsection{Gauge sector}
\label{sec:gaugesector}
In this paper, we work under the assumption of manifest left-right symmetry of the UV-Lagrangian, i.e., $g_R = g_L = g$. Here, $g_{L(R)}$ are the gauge couplings of $SU(2)_{L(R)}$, and $g$ is the $SU(2)_L$ gauge coupling of SM.
The mass matrix for charged gauge bosons is
\bea 
\mathcal{L}_{\text{mass}} \supset 
\frac{g^2}{8} \begin{pmatrix}
W_L^+ & W_R^+
\end{pmatrix}
\begin{pmatrix}
v^2 & -2 \k_1 \k_2 \\
-2  \k_1 \k_2  & V^2
\end{pmatrix}
\begin{pmatrix}
 W_L^- \\
 W_R^- \\
\end{pmatrix}   \,\,,
\eea
where, $v^2 = \k_1^2 + \k_2^2 + v_L^2$ and $V^2 = \k_1^2 + \k_2^2 + v_R^2$. The physical charged gauge bosons have masses,
\bea 
m^2_{W_{1,2}} = \frac{g^2}{4} \Big(v^2 +  V^2   \mp \sqrt{(v^2 -  V^2)^2 + 16  \k_1^2 \k_2^2 } \Big) \,\,\,,
\label{Wmasses}
\eea
$W_1^{\pm}$ is identified as the SM $W^{\pm}$ boson and $W_2^{\pm}$ is the new charged gauge boson with mass $\sim \mathcal{O}(v_R)$. The mixing matrix is characterized by an orthogonal rotation with angle $\xi  \simeq   2\k_1\k_2/v_R^2$. 

Similarly, the neutral gauge boson mass matrix is,
\bea 
\mathcal{L}_{\text{mass}} \supset \frac{1}{8}\begin{pmatrix}
W^{3\m}_{L} & W^{3\m}_{R} & B^{\m}
\end{pmatrix} 
\begin{pmatrix}
g^2 v^2 & -g^2 \k_+^2 & -g g_{BL} v_L^2 \\
 & g^2 V^2 & -g g_{BL} v_R^2 \\
 & & g_{BL}^2 (v_L^2 + v_R^2) 
\end{pmatrix}
\begin{pmatrix}
W^3_{L\m} \\ W^3_{R\m} \\ B_{\m} 
\end{pmatrix} \,\,, \nn\\
\eea
where $\k_+^2 = \k_1^2 + \k_2^2$, $g_{BL}$ is the gauge coupling of $U(1)_{B-L}$ and here some of the elements have been suppressed since the matrix is symmetric. The lightest eigenstate is massless and identified as the photon, while the other two states have masses
\bea
m^2_{Z_1,Z_2} &=& \frac{1}{8}\Big(g^2 v^2 + g^2 V^2 + g_{BL}^2 (v_L^2 + v_R^2) \nn\\
&&\mp \sqrt{(g^2 v^2 + g^2 V^2 + g_{BL}^2 (v_L^2 + v_R^2))^2 + 4 (g^4 + 2 g^2 g_{BL}^2)(\k_+^4 - v^2 V^2) } \Big) \,\,.\nn\\
\label{Zmasses}
\eea
The lighter mass eigenstate $Z_1$ corresponds to the SM $Z$ boson, while $Z_2$ has a mass  $\sim\mathcal{O}(v_R)$. 

In the limit $\k_1, \k_2, v_L \ll v_R$  the mixing matrix is~\cite{Dev:2016dja} 
\bea 
\begin{pmatrix} A_{\m} \\ Z_{1\mu}  \\ Z_{2\m} \end{pmatrix} =
\begin{pmatrix} 
s_W & c_W s_Y & c_W c_Y \\  -c_W & s_W s_Y & s_W c_Y \\ 0 & c_Y & s_Y
\end{pmatrix}
  \begin{pmatrix} W^3_{L\m} \\ W^3_{R\m} \\ B_{\m}  \end{pmatrix} ,
\eea 
where
\bea 
s_W &\equiv & \sin \theta_W = \frac{\gbl}{\sqrt{g^2 + 2 \gbl^2 }}\,\,,\,\,\,\,\,
c_W \equiv  \cos \theta_W = \sqrt{\frac{g^2 + \gbl^2}{g^2 + 2 \gbl^2}}\,\,,\nn\\
s_Y &\equiv &  \sin \theta_Y = \frac{\gbl}{\sqrt{g^2 + \gbl^2}}\,\,,\,\,\,\,\,
c_Y \equiv  \cos \theta_Y = \frac{g}{\sqrt{g^2 + \gbl^2}}\,\,. 
\eea
We fix $g_{BL} = g g'/(g^2 -g'^2)^{1/2}$, where $g'$ is the gauge coupling for $U(1)_Y$ of SM. Direct searches for spin-1 resonances have put a lower limit on the masses of the new charged and neutral gauge bosons. In DLRSM, the masses of such new gauge bosons are $m_{W_2} \sim g v_R/2 = $ and $m_{Z_2} \sim m_{W_2}/\cos \theta_Y$. Recently, the  lower limit on the mass of $W_2$ in DLRSM has been estimated to be, $m_{W_2} > 5.1$\,TeV\,\cite{Solera:2023kwt}, which leads to a lower bound on $v_R$, $v_R > 2 m_{W_2}/g = 15.7$\,TeV. The constraint on $m_{Z_2}$ is comparatively weaker, $m_{Z_2}>4.3$ TeV. Therefore, the lowest value of $v_R$ we use in our benchmark scenarios is $v_R = 20$\,TeV.

\subsection{Theoretical bounds}\label{subsec: theory}
We incorporate the following theoretical constraints:
\begin{itemize}
    \item {\it Perturbativity:} The quartic couplings of the scalar potential, $\{\l_{1,2,3,4},~\a_{1,2,3,4},~\rho_{1,2}\}$, are subjected to the upper limit of $4\pi$ from perturbativity. Moreover, the Yukawa couplings of the DLRSM Lagrangian must satisfy the perturbativity bound $y_{33}, \tilde{y}_{33} < \sqrt{4\pi}$, with $y_{33}, \tilde{y}_{33}$  defined in eq.\,\eqref{eq: y_33},
    These constrain the value of \vev ratios roughly to $r < 0.8$ and $w < 3.5$ \cite{Karmakar:2022iip}. 
    
    \item {\it Unitarity:} The scattering amplitudes of $2 \rightarrow 2$ processes involving scalars and gauge bosons must satisfy perturbative unitarity. To $\mathcal{O}(\k_1/v_R)$, these constraints can be expressed in terms of the masses of the new scalars in DLRSM \cite{Bernard:2020cyi},
    \bea
&0 <   \rho_1  <  \frac{8 \pi}{3}\, , \,\,\text{or,}\,\, \frac{m_{H_3}^2}{v_R^2}  < \frac{16 \pi}{3} \,\,\,,&  \nn\\ 
&\frac{(c_{H_3})^2}{k^4} \, \frac{m^2_{H_3}}{v_R^2} < \frac{16 \pi}{3} \,\,\, ,& \nn\\
&2 \frac{w^2}{k^2} \sum_{i=1,2} F_i^2 \frac{m^2_{H^\pm_i}}{v_R^2} + \frac{c_{H_3}}{k^2} \frac{m^2_{H_3}}{v_R^2}  <  16 \pi\,\,\,,& \nn\\
&2 \frac{w^2}{k^2} \sum_{i=1,2} S_i^2 \frac{m^2_{H^\pm_i}}{v_R^2} + \frac{c_{H_3}}{k^2} \frac{m^2_{H_3}}{v_R^2}  <  16 \pi\,\,\, ,&
\label{eq:unitaritycondition}
\eea
    where $k^2= 1+r^2+w^2$ and $F_i, S_i,$ and $c_{H_3}$ are defined in terms of the parameters of the potential \cite{Bernard:2020cyi}.
\item {\it Boundedness from below:} The scalar potential must be bounded from below\,(BFB) along all directions in field space. This leads to additional constraints on the quartic couplings of the model. The full set of such constraints was derived in ref.\,\cite{Karmakar:2022iip}, which we have implemented in our numerical analysis.   
\end{itemize}

\subsection{Constraints from $h(125)$ data}\label{subsec: higgs}
\label{sec: hdata_constraints}

In the following, we qualitatively describe the constraints on DLRSM from Higgs-related measurements at the LHC. 

\begin{itemize}
    \item  The key constraint comes from the measurement of the mass of SM-like Higgs, $m_h = 125.38 \pm 0.14\,$GeV \cite{CMS:2012qbp}. If the theoretical bounds of perturbativity and boundedness from below are taken into account together with $m_{h, \text{analytic}} \simeq 125$\,GeV, it leads to an upper bound on the \vev ratio, $w \lesssim 2.93 + 4.35 r - 0.48 r^2$. 

    \item One of the most stringent constraints on the DLRSM parameter space comes from the measurement of $hb\bar{b}$ coupling, $\k_b = 0.98^{+0.14}_{-0.13}$\,\cite{ATLAS:2020qdt}. If the mixing between $\phi^0_{1r}$ and $\phi^0_{2r}$ takes large values, $\k_b$ can deviate from unity, thereby ruling out a large region of parameter space allowed by theoretical bounds and the measurement of $m_h$. However, $ht\bar{t}$ coupling is not significantly modified and does not result in any new constraints. 
    
    \item As discussed in Sec.~\ref{sec:gaugesector}, a large value of $v_R$ ensures that the mixings between the SM-like and heavier gauge bosons are rather small, $\xi \sim \mathcal{O}(v^2/v_R^2)$. Therefore, the $h W_1 W_1$ and $h Z_1 Z_1$ couplings are quite close to their SM values and do not lead to any additional constraints on the DLRSM parameter space.
        
    \item The trilinear coupling of the SM-like Higgs given in eq.\,\eqref{eq: chhh}, does not necessarily align with the SM value. As seen in eq.\,\eqref{eq: chhh}, for non-zero mixings, particularly, $O_{21}\neq 0$, the parameters appearing in the paranthesis can individually take a wide range of values, leading to a potentially significant deviation of $c_{h^3}$ from $c^{\rm{SM}}_{h^3}$. In our analysis, we impose the ATLAS bound of $\k_h = [-2.3, 10.3]$ at 95$\%$ CL\,\cite{ATLAS:2019pbo}. 
\end{itemize}

\section{Effective Potential}\label{sec: effective potential}
In this section, we construct the full one-loop finite temperature effective potential\,\cite{Quiros:1999jp,Laine:2016hma} required to study the nature of the PT associated with the breaking of $SU(2)_R \times U(1)_{B-L}$. Below we describe the procedure step by step. 

The tree-level effective potential is obtained by setting all the fields to their respective background field value in the potential given in eq.\,\eqref{eq: potential}. The CP-even neutral component of $\chi_{\rm{R}}$ is responsible for breaking the $SU(2)_{\rm{R}}\times U(1)_{B-L}$ gauge group, whose background value we denote by $R$. Since $v_R\gg v$, all other field values can be set to zero. 
Hence, in the notation of eq.\,\eqref{eq: varphi}, the background fields are
$$\langle\varphi\rangle = \{0,0,0,R,0,0,0,0,0,0,0,0\}.$$
The tree-level effective potential is then given by
\beq
    V_0(R) = -\frac{\mu_3^2}{2}R^2 + \frac{\rho_1}{4}R^4.
\eeq
At the one-loop level, the zero-temperature correction to the effective potential is given by the Coleman-Weinberg (CW) formula \cite{ColWein}. In the Landau gauge, with $\overline{\rm{MS}}$ renormalization scheme, the CW potential is\,\cite{Quiros:1999jp}
\beq
    V_{\cw}(R) = \frac{1}{64\pi^2}\sum_i (-1)^{f_i} n_i m_i^4(R)\left[\log\bigg(\frac{m_i^2(R)}{\mu^2}\bigg) - c_i \right],
\eeq
where $i$ runs over all species coupling to the $SU(2)_R\times U(1)_{B-L}$-breaking field $\chi^0_{Rr}$. The field-dependent mass, $m_i(R)$ is the mass of the species $i$ in the presence of the background field $R$. When there is mixing between the different species, the masses are extracted as the eigenvalues of the corresponding mass matrices. The expressions for the field-dependent masses can be found in Appendix\,\ref{appendix: field}. In Appendix\,\ref{sec: neutrino mass} we take the minimal mechanism of neutrino mass generation of refs.\,\cite{Babu:1988qv, FileviezPerez:2016erl} and show that the right-handed neutrino $\n_R$ and the extra charged scalar do not contribute to the effective potential. Therefore the contributions only come from the CP-even scalars: $\{\phi_{1r}^0, \phi_{2r}^0, \chi_{Lr}^0, \chi_{Rr}^0\}$, CP-odd scalars: $\{\phi_{1i}^0, \phi_{2i}^0, \chi_{Li}^0, \chi_{Ri}^0\}$, charged scalars: $\{\phi_1^{\pm}, \phi_2^{\pm}, \chi_L^{\pm}, \chi_R^{\pm}\}$, and gauge bosons $W_{L,R}^{\pm}$, $Z_{L,R}$ and $B$. The factor $f_i$ is 0 (1) for bosons\,(fermions), and the number of degrees of freedom $n_i$ are,
\bea
&n_{\phi^0_{1r}}=n_{\phi^0_{2r}}=n_{\chi^0_{Lr}}=n_{\chi^0_{Rr}}= 1,& \nn\\
&n_{\phi^0_{1i}}=n_{\phi^0_{2i}}=n_{\chi^0_{Li}}=n_{\chi^0_{Ri}}=1,& \nn\\
&n_{\phi^{\pm}_{1}}=n_{\phi^{\pm}_{2}}=n_{\chi^{\pm}_{L}}=n_{\chi^{\pm}_{R}}=2,&\nn\\
&n_{W^{\pm}_{Lt}}=n_{W^{\pm}_{Rt}}=4,&\nn\\
&n_{W^{\pm}_{Ll}}=n_{W^{\pm}_{Rl}}=2,&\nn\\
&n_{Z_{Lt}}=n_{Z_{Rt}}=n_{B_{t}}=2,&\nn\\
&n_{Z_{Ll}}=n_{Z_{Rl}}=n_{B_{l}}=1.&\nn
\eea
The subscripts $t$ and $l$ stand for transverse and longitudinal polarizations of the gauge bosons. The constant $c_i = 5/6$ for gauge bosons, and $3/2$ for all other fields. We set the renormalization scale $\mu = v_R$ to ensure the validity of the CW formula by having $\ordone$ logs.

We impose the `on-shell' renormalization condition so that the position of the minimum and the mass of the CP-even scalar $\chi^0_{Rr}$ calculated from the one-loop potential coincides with the corresponding tree-level value. This is achieved by introducing a counter-term potential
\beq
    V_{\rm{c.t.}}(R) = -\frac{\delta\mu_3^2}{2} R^2 + \frac{\delta\rho_1}{4}R^4\,,
\eeq
where the unknown coefficients $\delta\mu_3^2$ and $\delta\rho_1$ are fixed by demanding
\begin{subequations}
\beq
     \left. \frac{\partial}{\partial R}(V_{\cw} + V_{\rm{c.t.}})\right\vert_{R=v_R} = 0\,,
\eeq
\beq
     \left. \frac{\partial^2}{\partial R^2}(V_{\cw} + V_{\rm{c.t.}})\right\vert_{R=v_R} = 0\,.
\eeq
\end{subequations}
This leads to
\begin{subequations}
\beq
     \delta\mu_3^2 = \frac{3}{2 v_R}\left.\frac{\partial V_{\cw}}{\partial R}\right\vert_{R=v_R} - \frac{1}{2}\left.\frac{\partial^2 V_{\cw}}{\partial R^2}\right\vert_{R=v_R},
\eeq
\beq
     \delta\rho_1 = \frac{1}{2 v_R^3}\left.\frac{\partial V_{\cw}}{\partial R}\right\vert_{R=v_R} - \frac{1}{2 v_R^2}\left.\frac{\partial^2 V_{\cw}}{\partial R^2}\right\vert_{R=v_R}.
\eeq
\end{subequations}
Then the one-loop contribution to the effective potential is
\begin{equation}
    V_1 = V_{\cw}+V_{\rm{c.t.}}.
\end{equation}

Next, we include the one-loop finite temperature correction \cite{Quiros:1999jp, Dolan:1973qd}
\beq\label{eq: V1T}
    V_{1T}(R,T) = \frac{T^4}{2\pi^2}\sum_i (-1)^{f_i} n_i J_{b/f}\left(\frac{m_i^2}{T^2}\right),
\eeq
where the functions $J_{b/f}$ are given by
\beq\label{eq: Jbf}
  J_{b/f}(x^2) = \int_0^{\infty} dy~ y^2 \log[1\mp e^{-\sqrt{y^2+x^2}}]\,.
\eeq
In the high-T approximation, i.e. $x^2\equiv\frac{m_i^2}{T^2}\ll 1$, eq.\,\eqref{eq: Jbf} simplifies to\,\cite{Cline_1997}
\bea
J_f(x^2) \approx &-&\frac{7\pi^4}{360} + \frac{\pi^2}{24} x^2 + \mathcal{O}(x^4)\,,\label{eq: highTf}\nn\\
J_b(x^2) \approx &-&\frac{\pi^4}{45} + \frac{\pi^2}{12} x^2 - \frac{\pi}{6}(x^2)^{3/2}+\mathcal{O}(x^4)\,.\label{eq: highTb}
\eea
The non-analytic $(x^2)^{3/2}$ term present in the bosonic case is mainly responsible for the formation of a barrier between the minima of the effective potential at zero and non-zero field values, leading to a FOPT. 

In addition to the one-loop terms, multi-loop contributions from daisy diagrams need to be re-summed to cure the infrared divergence arising from the bosonic zero-modes \cite{Carrington:1991hz}. There are two ways to do this: the Parwani method \cite{Parwani:1991gq} and the Arnold-Espinosa method \cite{Arnold-Espinosa}. In the Parwani method, the field-dependent mass is replaced with thermally corrected mass, i.e., $m_i^2(R)\rightarrow m_i^2(R) + \Pi_i(T)$, in the expressions of $V_{\cw}$ and $V_{1T}$. Here $\Pi_i$ is the thermal mass obtained using the high-T expansion of $V_{1T}$, as shown in Appendix\,\ref{appendix: field}. The daisy re-summed effective potential is given by
\beq
    V_{\rm{eff}} = V_0 + V_{\cw}(m_i^2(R) + \Pi_i(T)) + V_{\rm{c.t.}} + V_{1T}(m_i^2(R) + \Pi_i(T))\,.
\eeq

In the Arnold-Espinosa method, no such replacement for field-dependent mass is made, but an extra daisy term is added to the effective potential: 
\beq\label{eq: Vdaisy}
    V_{\rm{D}} = -\frac{T}{12\pi}\sum_i n_i \bigg((m_i^2(R) + \Pi_i(T))^{3/2} - (m_i^2(R))^{3/2}\bigg)\,.
\eeq
Thus the effective potential is given by
\beq
    V_{\rm{eff}} = V_0 + V_{\cw} + V_{\rm{c.t.}} + V_{1T} + V_{\rm{D}}.
\eeq
In our analysis, we use the Arnold-Espinosa method, as it takes into account the daisy resummation consistently at the one-loop level, while the Parwani method mixes higher-order loop effects in the one-loop analysis. 

\begin{figure}[tbp] 
\centering 
\includegraphics[width=.49\textwidth]{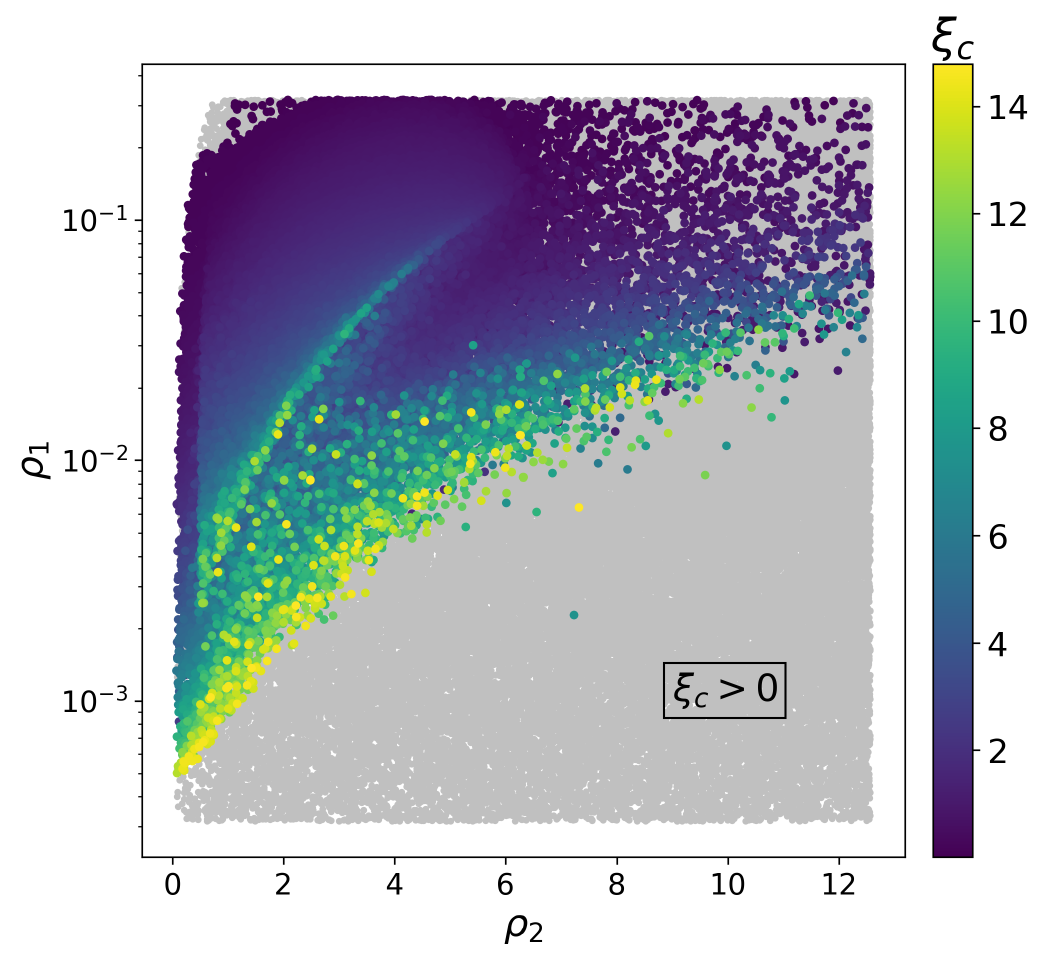}
\includegraphics[width=.49\textwidth]{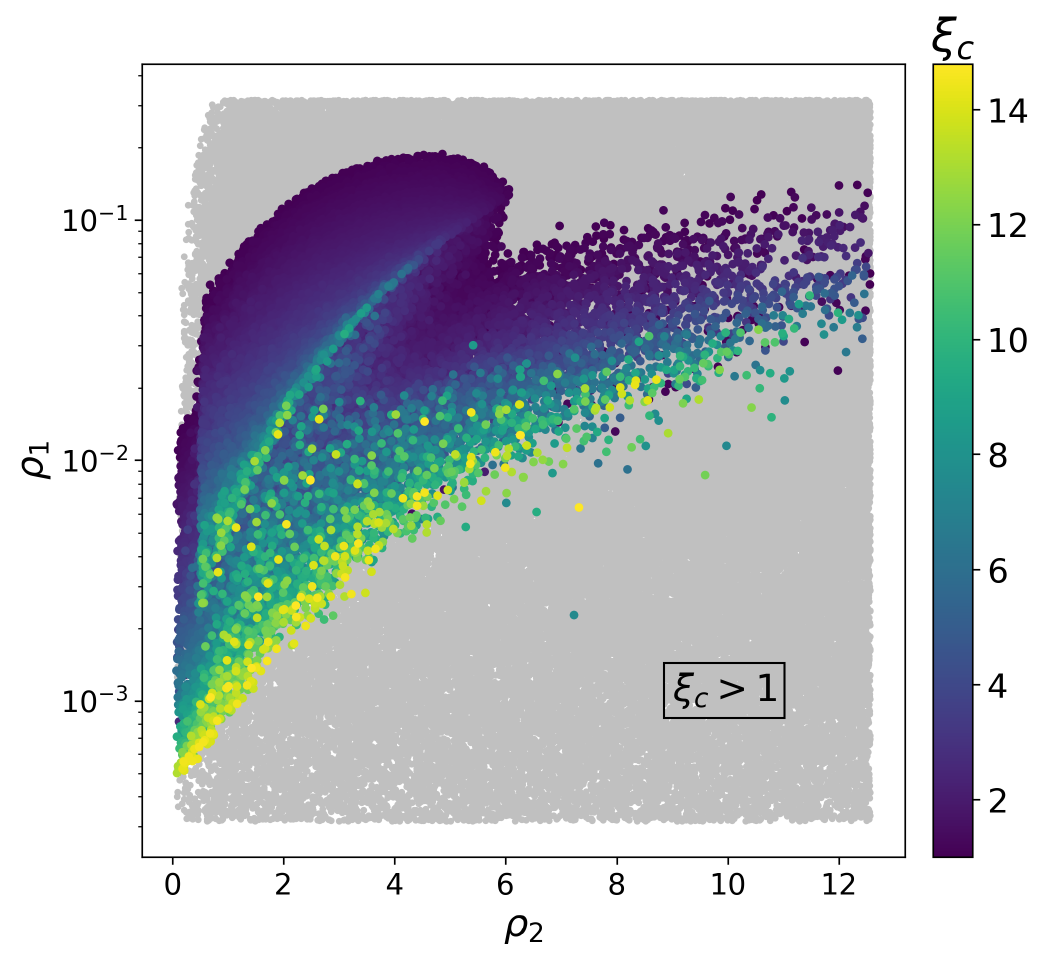}
\caption{\label{fig: rho1rho2} Points with strong FOPT in the $\rho_1-\rho_2$ plane, for $v_R=30$ TeV. The grey points have passed the theoretical and experimental constraints. The left panel shows all points showing FOPT with $\xi_c>0$, whereas in the right panel, the points satisfy the condition $\xi_c>1$.}
\end{figure}

\section{Parameter scan}
\label{sec: paramscan}
As discussed earlier, DLRSM has a large number of parameters: ten quartic couplings, along with $r,w,\mu_4$, and $v_R$. This is called the \textit{generic basis}. To reduce the number of parameters for our analysis, we work in the \textit{simple basis}, introduced in ref.\,\cite{Karmakar:2022iip}. The condition of boundedness from below, discussed in Sec.\,\ref{subsec: theory}, requires that the ratio $x = \l_2/\l_4$ is restricted to the range $x \in [0.25, 0.85]$. Therefore, we keep $\l_2$ as a separate parameter, while we equate $\l_1=\l_3=\l_4\equiv \l_0$. Similarly, guided by the approximate mass relation, $m_{H_1}\approx \frac{1}{2}(\a_3-\a_4)$, we allow for the possibility of having $\a_3\neq \a_4$ by keeping them independent, while setting $\a_1=\a_2=\a_4\equiv\a_0$. Thus the {\it simple basis} contains six quartic couplings
\bea 
&&\{ \l_0,~\l_2,~\a_0,~\a_3,~\rho_1,~\rho_2 \}\,.
\eea
Along with these quartic couplings, we also scan over the \vev ratios $r$, $w$, and take $v_R = 20,~30,$ 50 TeV. As the mass parameter $\mu_4$ plays an insignificant role in the effective potential, we set $\mu_4=0$ in our analysis. Using the {\it simple basis} allows us to capture the key features of GW phenomenology of DLRSM while retaining the interplay of the existing theoretical and collider constraints.

In preliminary scans, we find that promising scenarios of strong first-order phase transition occur for small values of $\rho_1$. For points with relatively large couplings, the daisy potential, $V_{\rm{D}}$, given in eq.\,\eqref{eq: Vdaisy} starts dominating over the contribution from the thermal potential, $V_{1T}$, given in eq.\,\eqref{eq: V1T}. When this happens, the symmetry-restoring property of the finite temperature effective potential is lost and instead, symmetry non-restoration is observed. Then the minimum at the non-zero field value becomes deeper at high temperatures, implying the absence of a phase transition, as discussed in refs.\,\cite{Weinberg:1974hy,Kilic:2015joa,Meade:2018saz}. Based on these observations, we choose the following parameter ranges: 
\bea
    &\log\a_0 \in [-3,0],~\log\a_3 \in [-3,0],~\log\rho_1 \in [-3.5,-0.5],~\rho_2 \in [0,4\pi],&\nn\\
    &~x \in [0.25, 0.85],~\log r \in [-3,0],~\log w \in [-6,1],~v_R = 20, 30, 50\,\text{TeV}.&
\eea
Each parameter is selected randomly from a uniform distribution in the respective range. The parameter $\lambda_0$ is chosen in the following manner:
\begin{itemize}
\item To increase the number of points satisfying the bound on SM-like Higgs mass ($m_h$), we solve the equation, $m_{h, \text{analytic}}\,(\l_0 = \Lambda_0) = 125.38$\,GeV, for a fixed set of values $\{\a_0, \a_3, \rho_1, \rho_2, x\}$. 
\item Using the solution $\Lambda_0$, we choose a random value of $\l_0$ as, $\lambda_{0} = (1+y)\,\Lambda_0$, with $y \in [-0.1, 0.1]$. 
\item Finally, each parameter point is defined by the set: $$\big\{\l_0,~\l_2 = x\l_0,~\a_0,~\a_3,~\rho_1,~\rho_2,~r,~w,~v_R\big\}$$
\end{itemize}
Given a parameter point, we first check if it satisfies the theoretical constraints: boundedness from below, perturbativity, and unitarity, discussed in Sec.\,\ref{subsec: theory}. Next, the Higgs constraints described in Sec.\,\ref{subsec: higgs} are checked. Furthermore, the constraint from meson mixing $m_{H_1} > 15$\,TeV is imposed.

If the parameter point passes all the aforementioned theoretical and experimental constraints, we construct the effective potential using the Arnold-Espinosa method. We satisfy the Linde-Weinberg bound\,\cite{Linde:1975sw,Weinberg:1976pe} by numerically checking that the minimum of the zero-temperature effective potential at $R=v_R$ is the absolute minimum. We reject the point if symmetry non-restoration persists at high temperatures. Next, we check for a possible first-order phase transition, using the python-based package \texttt{CosmoTransitions}\,\cite{Wainwright:2011kj}. The strength of FOPT can be quantified by the ratio
\beq
    \xi_c = \frac{v_c}{T_c},
\eeq
where $T_c$ is the critical temperature at which the two minima become degenerate and $v_c$ is the \vev at $T_c$. The FOPT is considered to be strong if the following criterion is met\,\cite{Quiros:1994dr} \footnote{It is known that the field value, $v_c$, and $T_c$ are gauge dependent, therefore, so is the ratio $\frac{v_c}{T_c}$. However changing the gauge-fixing parameter has a subleading effect on $\frac{v_c}{T_c}$\,\cite{Patel:2011th,Chatterjee:2022pxf}. In the subsequent analysis, we work in the Landau gauge, for which the gauge-dependence is numerically minimised. },
\beq
    \xi_c >1.
\eeq

\begin{figure}[tbp]
\centering 
\includegraphics[width=.98\textwidth]{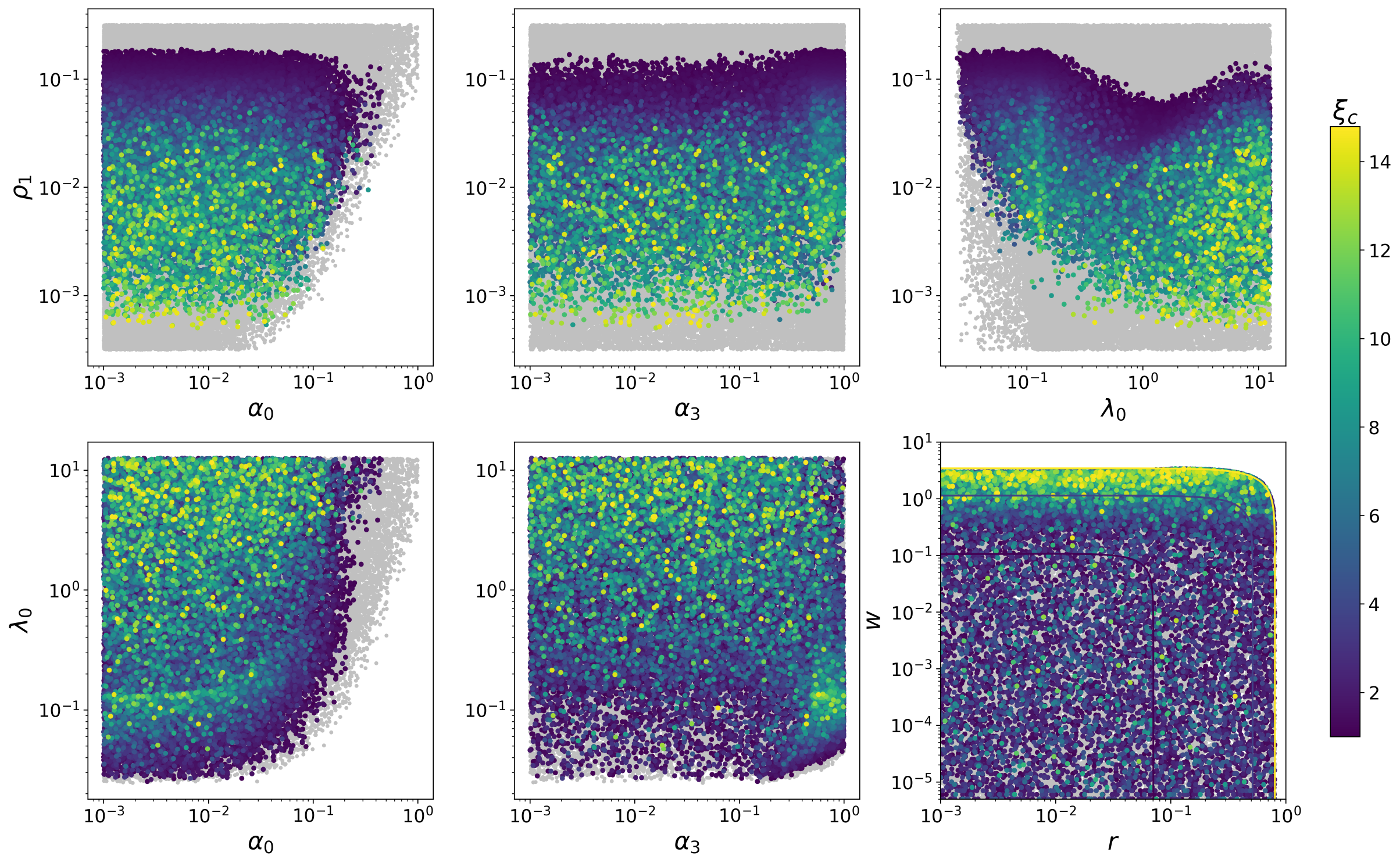}
\caption{\label{fig: projections} Projections showing the points with SFOPT on different parameter planes, for $v_R=30$ TeV. The grey points show all points passing the theoretical and experimental constraints. The points satisfy the condition $\xi_c>1$.}
\end{figure}

In fig.\,\ref{fig: rho1rho2}, we show the points with FOPT projected onto the $\rho_1-\rho_2$ plane for $v_R=30$ TeV, color-coded according to the value of $\xi_c$. The left panel shows all points with $\xi_c>0$, while the right panel only shows points satisfying the SFOPT criterion $\xi_c>1$. The grey dots depict parameter points passing the existing theoretical and experimental bounds. As suggested by the preliminary scans, SFOPT prefers $\rho_1 \lesssim \mathcal{O}(0.1)$. Points with $\rho_1\lesssim\mathcal{O}(10^{-2})$ and $\rho_2\gtrsim \ordone$ violate the Linde-Weinberg bound. Therefore, there are no points showing SFOPT in this region. A large number of points with $\rho_2 \gtrsim 6$ also exhibit symmetry non-restoration at high temperatures. 

Fig.\,\ref{fig: projections} shows various two-dimensional projections of the DLRSM parameter space for $v_R=30$\,TeV, depicting points with SFOPT. The parameter $\a_0$ is always smaller than 1, as indicated by the left panels in the top and the bottom row. We also restrict ourselves to $\a_3<1$ to avoid points showing symmetry non-restoration. Along the $\a_3$ direction, there is a sharp change in the density of points around $\a_3\approx 0.5$, coming from the bound $m_{H_1} > 15$\,TeV. The value of $\a_3$ where the density changes is different for $v_R = 20$, 30, and 50 TeV. Since the couplings are small for a large number of parameter points, the approximate relation given in eq.\,\eqref{eq: mH_approx} tells us that $\l_1$ can take values close to $\l_{\rm{SM}}\approx 0.13$. In the top right and bottom left panels, we indeed observe an over-density of points clustered around $\l_0\approx 0.13$. In the $\rho_1-\l_0$ plane, a majority of points with large $\xi_c$ occur for small $\rho_1$, and large $\l_0$. In the $r-w$ plane, points with large $\xi_c$ occur mostly at higher values of $w$ ($\gtrsim \mathcal{O}(0.1)$) and are less frequent for smaller values of $w$. So this parameter region can lead to a detectable GW background. There is no preference along the $r$ direction. The points with large $\xi_c$ also have relatively large values of $y_{33}$, as indicated by the contours corresponding to $y_{33}=1,~1.5$, and $\sqrt{4\pi}$.

The strength of FOPT is more rigorously characterized by three parameters, $\a,~\beta/H_*,$ and $T_n$, which are required to compute the GW spectrum. These are defined as follows:
\begin{itemize}
    \item The probability of tunneling from the metastable to the stable minimum is given by\,\cite{Linde:1980tt}
    \beq
        \Gamma (T) \approx T^4 \left(\frac{S_3}{2\pi T}\right)^{3/2} e^{-\frac{S_3}{T}},
    \eeq
    where $S_3$ is the $O(3)$-symmetric Euclidean bounce action. This is calculated using the tunneling solution of the equation of motion of the scalar field. We use \texttt{CosmoTransitions}\,\cite{Wainwright:2011kj} to compute $S_3$. The probability of nucleating a bubble within a Hubble volume increases as the universe cools below $T_c$, and  becomes $\ordone$ at the nucleation temperature, $T_n$. This happens when 
\beq
\Gamma(T_n)\approx \big(H(T_n)\big)^4.
\eeq
In the radiation-dominated era, this implies\,\cite{Huang:2020bbe}
\beq\label{eq: nuclCriterion}
\frac{S_3(T_n)}{T_n} \simeq -4 \ln\left(\frac{T_n}{m_{\rm{Pl}}}\right),
\eeq
where the Planck mass $m_{\rm{Pl}} = 1.22\times 10^{19}$ GeV. 
    \item The parameter $\a$ is the vacuum energy released during the transition, $\rho_{\rm{vac}}$, normalized by the radiation density at the time of FOPT\,\cite{Espinosa:2010hh},
    \beq\label{eq: alpha}
        \alpha \equiv \frac{\rho_{\rm{vac}}}{\rho_{\rm{rad}}}, 
    \eeq
    where,
    \bea
        \rho_{\rm{vac}} &=& (\left.V_{\rm{High}} - V_{\rm{Low}}) - \frac{T}{4}\bigg(\frac{\partial V_{\rm{High}}}{\partial T}-\frac{\partial V_{\rm{Low}}}{\partial T}\bigg) \right\vert_{T = T_*}\,,\\
        \rho_{\rm{rad}} &=& \frac{\pi^2}{30} g_* T_*^4\,.
    \eea
    Here $T_*$ is the temperature of the universe at the time when dominant GW production takes place. We take $T_* \simeq T_n$ in our calculations. The subscripts `High' and `Low' refer to the metastable and stable minima respectively, at the time of tunneling. $g_*$ is the number of relativistic degrees of freedom at $T=T_*$. For DLRSM, $g_* = 130$.
    \item $\beta$ is related to the rate or inverse duration of the phase transition, defined as\,\cite{Caprini:2015zlo}  
    \beq\label{eq: beta}
\beta \equiv  -\left.\frac{dS}{dt}\right\vert_{t=t_*} = TH_*\left.\frac{dS}{dT}\right\vert_{T=T_*},
\eeq
where, $S=S_3/T$ and $H_*$ is the Hubble's constant at $T=T*$. 
\end{itemize}

 \begin{figure}[tbp]
\centering 
\includegraphics[width=.99\textwidth]{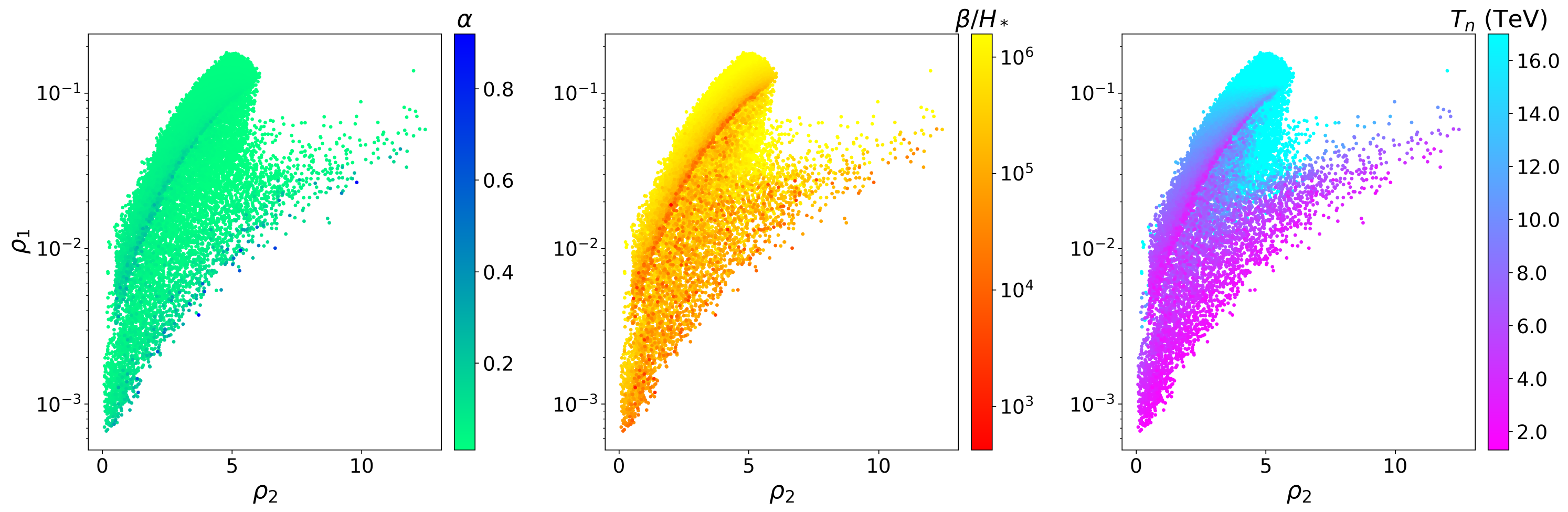}
\caption{\label{fig: abt} Variation of PT parameters in the $\rho_1-\rho_2$ plane. Color code shows the variation of $\alpha$ (left panel), $\beta/H_*$ (middle panel), and $T_n$ (right panel). Here, $v_R=30$ TeV.}
\end{figure}

 For points satisfying $\xi_c>1$, we compute the nucleation temperature $T_n$. We find the solution of eq.\,\eqref{eq: nuclCriterion} using the secant method, where the tunneling action is calculated by \texttt{CosmoTransitions}. We remove any points with $T_n<0$, as this indicates that the PT is not completed till the present time. Moreover, we set a lower bound of $T_n > 500\,$GeV to ensure that the PT is completed before the EW epoch. Once $T_n$ is obtained, $\a$ and $\beta/H_*$ can be computed using eqs.\,\eqref{eq: alpha} and \eqref{eq: beta} respectively. Fig.\,\ref{fig: abt} shows the variation of the PT parameters $\a$ (left panel), $\beta/H_*$ (middle panel), and $T_n$ (right panel), in the $\rho_1-\rho_2$ plane. The evaluated ranges roughly are, $\a\in[0,0.8]$, $\beta/H_*\in[10^2,10^6]$, and $T_n\in [2,16]$ TeV. $T_n$ is observed to take smaller values in regions where the strength of SFOPT is high. 

\section{Gravitational wave background}\label{sec: GW}
The GW spectrum is defined as \cite{Caprini:2015zlo}
\beq
\Omega_{\rm{GW}}(f)\equiv \frac{1}{\rho_c}\frac{d\rho_{\rm {GW}}}{d\ln f},
\eeq
where $f$ is the frequency, $\rho_{\rm {GW}}$ is GW energy density, and $\rho_c$ is the critical energy density of the universe, given by,
\beq
\rho_c = \frac{3H_0^2}{8\pi G}.
\eeq
\noindent Here, $H_0 = 100\,h~{\rm{km \,s^{-1} Mpc^{-1}}}$ is the Hubble constant with the current value of $h=0.6736\pm 0.0054$\,\cite{Planck:2018vyg} and $G$ is  Newton's gravitational constant. 

A strong FOPT proceeds by nucleation of bubbles of the stable phase which expand rapidly in the sea of the metastable phase. GWs are produced when the expanding bubbles collide and coalesce with each other. If sufficient friction exists in the plasma, the bubble walls may reach a terminal velocity $v_w$. We take $v_w = 1$ in our analysis. GW production happens via three main processes: bubble wall collisions ($\Omega_{\rm{col}}$), sound waves produced in the thermal plasma ($\Omega_{\rm{sw}}$), and the resulting MHD turbulence ($\Omega_{\rm{turb}}$). For a recent review of the different GW production mechanisms, please refer to \cite{Athron:2023xlk}. In the non-runaway scenario \cite{Caprini:2015zlo}, GW production happens primarily through sound waves and turbulence, i.e.,
\beq
h^2 \Omega_{\rm{GW}} \simeq h^2 \Omega_{\rm{sw}}+h^2 \Omega_{\rm{turb}}\, ,
\eeq
where\,\cite{Guo:2020grp,Caprini:2015zlo},
\bea
h^2\Omega_{\rm{sw}}(f) &=& 2.65\times 10^{-6} \left(\frac{100}{g_*}\right)^{1/3}\left(\frac{H_*}{\beta}\right)^2 \left(\frac{\k_{\text{sw}}\a}{1+\a}\right)^2v_w ~S_{\rm{sw}}(f)~\Upsilon(\tau_{\text{sw}}),\label{eq: soundwaves}\\
h^2\Omega_{\rm{turb}}(f) &= & 3.35\times 10^{-4} \left(\frac{100}{g_*}\right)^{1/3}\left(\frac{H_*}{\beta}\right)^2 \left(\frac{\k_{\rm{turb}}\a}{1+\a}\right)^{3/2}v_w ~S_{\rm{turb}}(f)\,. \label{eq: turbulence}
\eea

\noindent Here, $\k_{\text{sw}}$ and $\k_{\text{turb}}$ are the efficiency factors for the respective processes. The efficiency factor $\k_{\text{sw}}$ is given by
\beq
\k_{\text{sw}} = \frac{\a}{0.73+0.083\sqrt{\a}+\a},
\eeq
and $\k_{\text{turb}}$ is known to be at most $5-10\%$ of $\k_{\text{sw}}$. Here we take $\k_{\text{turb}} = 0.05 ~\k_{\text{sw}}$. We have included the suppression factor $\Upsilon(\tau_{\text{sw}})$ that arises due to the finite lifetime $\tau_{\text{sw}}$ of sound waves \cite{Guo:2020grp},
\beq
\Upsilon(\tau_{\text{sw}}) = 1 - \frac{1}{1+2\tau_{\text{sw}}H_*}\,,
\eeq
with
\beq
\tau_{\text{sw}} = \frac{R_*}{\overline{U_f}}\,,
\eeq
where the mean bubble separation $R_*\simeq (8\pi)^{1/3} v_w/\beta$ and the mean square velocity is
\beq
\overline{U_f}^2 = \frac{3}{4}\frac{\a}{1+\a}\k_{\rm{sw}}\,.
\eeq

The spectral shape functions, $S_{\rm{sw}}$ and $S_{\rm{turb}}$ determine the behavior of each contribution at low and high frequencies. These are
\bea
S_{\rm{sw}}(f) &=& \left(\frac{f}{f_{\rm{sw}}}\right)^3 \left(\frac{7}{4+3(f/f_{\rm{sw}})^2}\right)^{7/2},\nn\\
S_{\rm{turb}}(f) &=& \left(\frac{f}{f_{\rm{turb}}}\right)^3\frac{1}{[1+(f/f_{\rm{turb}})]^{11/3}(1+8\pi f/h_*)}\,.
\eea

Here, $h_*$ is the Hubble rate at $T=T_*$, 
\begin{equation}
h_* = 1.65\times 10^{-7}~{\rm{Hz}}\left(\frac{T_*}{100~{\rm{GeV}}}\right)\left(\frac{g_*}{100}\right)^{1/6}.
\end{equation} 
The red-shifted peak frequencies, after taking into account the expansion of the universe, are,
\begin{eqnarray}
f_{\rm{sw}} &=& 1.9\times 10^{-5}{\rm{Hz}}\left(\frac{g_*}{100}\right)^{1/6}~\frac{1}{v_w}\left(\frac{\beta}{H_*}\right)\left(\frac{T_*}{100~{\rm{GeV}}}\right),\label{eq: fsw}\label{eq: f_sw}\\
f_{\rm{turb}} &=& 2.7\times 10^{-5}{\rm{Hz}}\left(\frac{g_*}{100}\right)^{1/6}~\frac{1}{v_w}\left(\frac{\beta}{H_*}\right)\left(\frac{T_*}{100~{\rm{GeV}}}\right).\label{eq: fturb}
\end{eqnarray}

\begin{figure}[tbp]
\centering
\includegraphics[width=.5\textwidth]{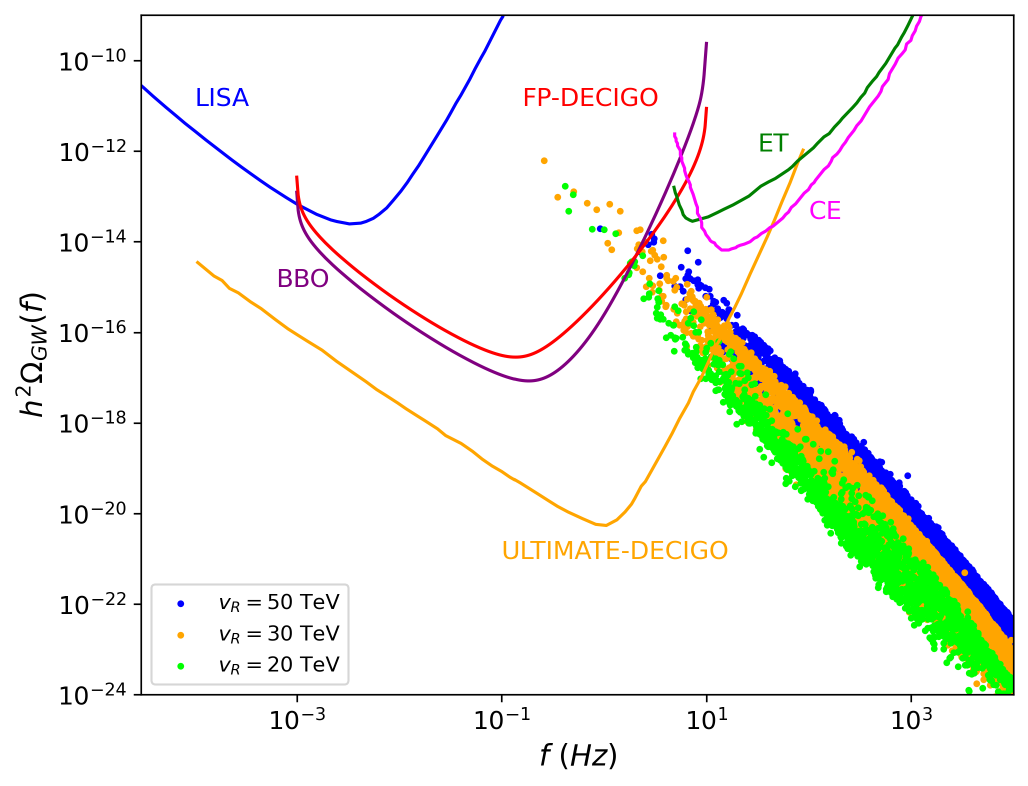}
\hfill
\includegraphics[width=.49\textwidth]{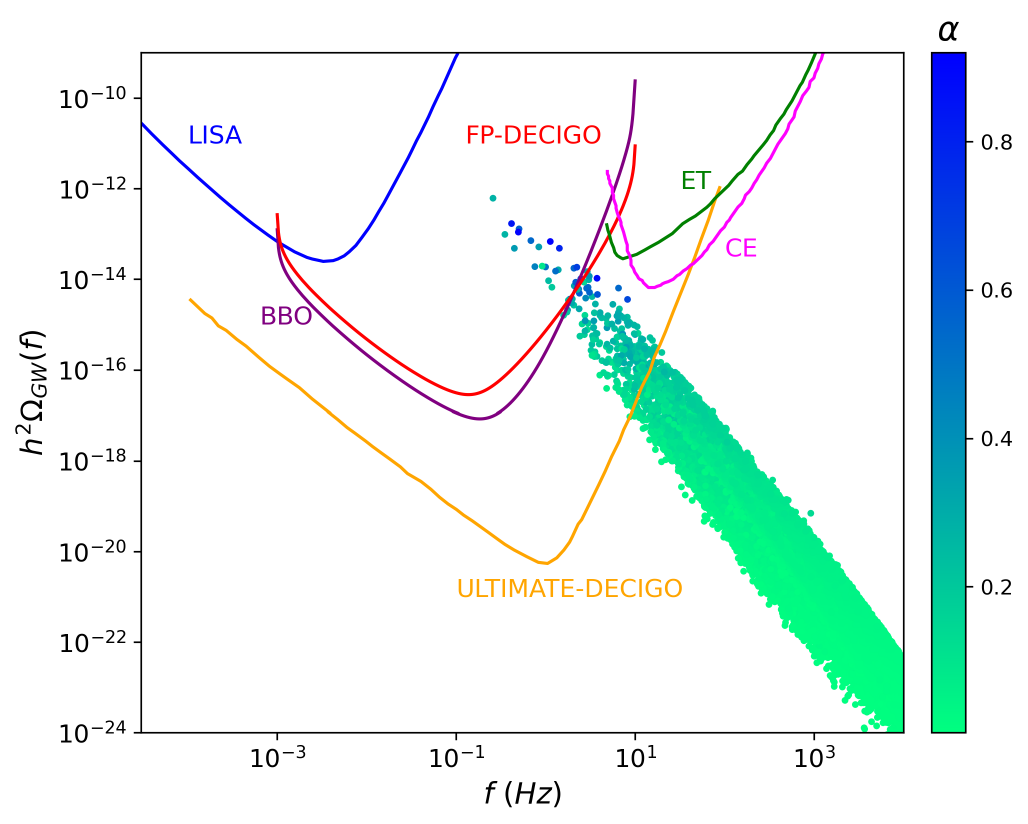}
\caption{\label{fig: gw_scatter} The peak of the GW spectrum $\Omega_{\rm{GW}}$ for points with SFOPT, along with the power-law integrated sensitivity curves of various upcoming GW detectors. Left panel: points corresponding to $v_R=20,~30$, and 50 TeV are shown. Right panel: Points are color-coded according to the value of $\a$, for $v_R=20,~30$, and 50 TeV combined.}
\end{figure}

From the expressions of $\Omega_{\rm{sw}}$ and $\Omega_{\rm{turb}}$, it is clear that large $\a$ and small $\beta/H_*$ lead to a strong GW spectrum. The peak frequency is proportional to $T_n\sim v_R$ and hence, the peak shifts to the right for larger $v_R$. This is illustrated in fig.\,\ref{fig: gw_scatter}, where we show scatter plots of the parameter points for which $\a,~ \beta/H_*$, and $T_n$ have been computed. Each point represents the peak value corresponding to the GW spectrum, $h^2\Omega_\text{GW}$. The left panel shows that these points shift to the right as $v_R$ is progressively increased between $v_R=20,~30$ and $50$ TeV. The strength of the GW signature is not affected by varying $v_R$. The right panel shows the variation of $\a$ for the points corresponding to $v_R = 20,~30$, and $50$ TeV combined. There is clearly a positive correlation between large $\a$ and the strength of GW. The solid lines represent the power-law integrated sensitivity curves\,\cite{Thrane:2013oya} corresponding to various planned detectors calculated for an observation time of $\tau = 1$ year, and a threshold SNR=1 (see eq.\,(\ref{eq: SNR})). The curve for Ultimate-DECIGO is obtained following the prescription of ref.\,\cite{Kuroyanagi:2014qza}, while the other curves are taken from\,\cite{schmitz_2020_3689582}. Points lying above the sensitivity curve of a detector feature SNR$>1$, and have strong detection prospects. The DLRSM phase transition has good detection prospects for the detectors FP-DECIGO, BBO, and Ultimate-DECIGO for the chosen set of $v_R$ values. The GW spectrum is too weak to be detected at ET and CE for the chosen range of $v_R$. If the scale $v_R$ is increased by a factor of $\sim 10-100$, these two detectors may be able to detect them, but we ignore this region as the complementary collider constraints would be too weak. 

\begin{figure}[tbp]
\centering 
\includegraphics[width=.99\textwidth]{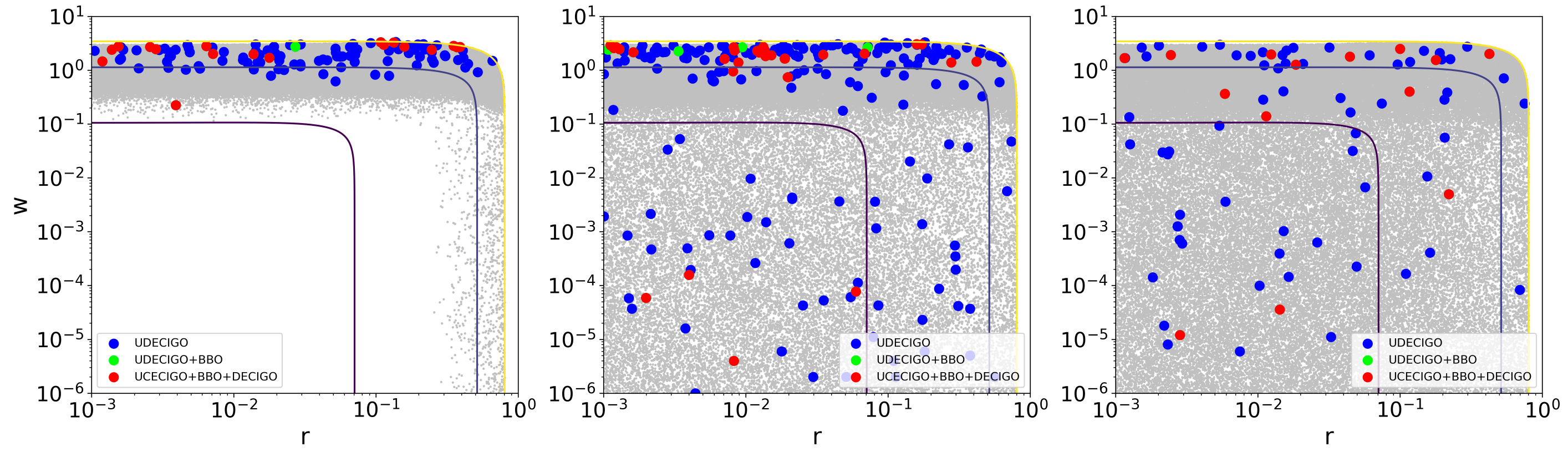}
\caption{\label{fig: rw_203050} Points with detectable GW signature at upcoming observatories: Ultimate-DECIGO (UDECIGO), BBO, and FP-DECIGO. The scale is chosen to be, $v_R = 20$ TeV (left panel), $v_R=30$ TeV (middle panel), and $v_R = 50$ TeV (right panel). The purple, blue, and yellow contours represent the upper limits on $y_{33} = 1, 1.5, \sqrt{4\pi}$ respectively based on eq.\,\eqref{eq: y_33}.}
\label{fig:rwplane}
\end{figure}

In fig.\,\ref{fig: rw_203050} we illustrate the distribution of the points with detectable GW signal in the $r-w$ plane. The grey points pass all the theoretical and experimental constraints. The blue points are only detectable at Ultimate-DECIGO, the green points are detectable by Ultimate-DECIGO as well as BBO, and the red points can be detected at all three detectors. Interestingly, for $v_R=20$ TeV, the red, green, and blue points are densely clustered around $w\sim\mathcal{O}(1)$. For most of these points, $y_{33}$ is also large, $y_{33} \sim 1.5 - \sqrt{4\pi}$. In the middle panel, i.e. $v_R=30$ TeV, the majority of points still prefer $w\sim\ordone$, but now there are also points at lower values of $w$. In the case of $v_R=50$ TeV, we see that the clustering of points around $\ordone$ values of $w$ is even more diffuse. In all three cases, i.e. $v_R = 20,~30,$ and 50 TeV, there is no particular preference in the $r$ direction, as also seen from the SFOPT plots given in fig.\,\ref{fig: projections}. 

\begin{table}[tbp] 
\begin{center}
\begin{tabular}{c | c c c c c c}
\hline
 & BP1   & BP2   & BP3    & BP4   & BP5  & BP6\\ 
\hline
$v_R$ (TeV) & 30        & 30       & 30       & 30       & 20         & 50       \\
$\lambda_0$ & 0.126796  & 0.466090 & 0.308396 & 0.324564 & 1.982649   & 0.799371 \\  
$\lambda_2$ & 0.097015  & 0.253725 & 0.141320 & 0.267655 & 1.670007   & 0.413236 \\ 
$\alpha_0$  & 0.004789  & 0.003504 & 0.007640 & 0.012450 & 0.012042   & 0.021020 \\  
$\alpha_3$  & 0.957421  & 0.005786 & 0.006466 & 0.004839 & 0.001015   & 0.003094 \\ 
$\rho_1$    & 0.019071  & 0.001274 & 0.001929 & 0.005930 & 0.009976   & 0.003445 \\  
$\rho_2$    & 2.003479  & 0.627225 & 1.166146 & 1.674371 & 5.574184   & 2.275937 \\
$r$         & 0.008261  & 0.008136 & 0.418869 & 0.020970 & 0.390416   & 0.424048 \\  
$w$         & $4\times 10^{-6}$  & 0.950364 & 1.439902 & 0.766492 & 2.702912   & 2.018973\\
\hline
$m_{W_R^{\pm}}$ (TeV) & 9.81  & 9.81   & 9.81     & 9.81     & 6.54      &  16.35  \\
$m_{Z_R}$ (TeV)       & 11.58 & 11.58  & 11.58    & 11.58    & 7.72      &   19.30  \\
$m_{H_1}$ (TeV)       & 20.72 & 15.97  & 32.79    & 20.89    & 81.06     & 99.90    \\
$m_{H_2}$ (TeV)       & 29.74 & 23.13  & 45.58    & 34.46    & 116.99    & 144.97 \\  
$m_{H_3}$ (TeV)       & 5.86  & 1.51   & 1.86     & 3.27     & 2.82      & 4.15 \\ 
\hline
$\alpha$    & 0.280     & 0.274    & 0.243    & 0.122    & 0.428      & 0.273 \\ 
$\beta/H_*$ & 422       & 1050     & 2648     & 8267     & 975        & 3204    \\  
$T_c$ (TeV) & 5.78      & 3.26     & 3.46     & 4.83     & 2.82       & 5.87  \\
$T_n$ (TeV) & 3.08      & 1.68     & 1.86     & 2.91     & 1.37       & 3.26  \\
\hline
\end{tabular}
\end{center}
\caption{\label{table: bp} Benchmark points for DLRSM in the simple basis.
}
\end{table}

\section{Detection prospects}\label{sec: detection prospects}
The prospect of detecting a GW signal in a given GW observatory can be quantified using the signal-to-noise ratio (SNR), defined as \cite{Athron:2023xlk,Thrane:2013oya}
\beq\label{eq: SNR}
    {\rm{SNR}} = \sqrt{n_{\rm{det}}\tau\int_{f_{\rm{min}}}^{f_{\rm{max}}} df \left[\frac{\Omega_{\rm{GW}}(f) h^2}{\Omega_{\rm{sens}}(f) h^2}\right]^2} ,
\eeq
where $\tau$ is the time period (in seconds) over which the detector is active, and the integration is carried out over the entire frequency range $[f_{\rm{min}},f_{\rm{max}}]$ of the detector. For calculations, we take $\tau = 1$ year. The factor $n_{\rm{det}}$ is two for experiments aimed at GW detection via cross-correlation measurement, or one for experiments aimed at detection via auto-correlation measurement (eg. LISA). $\Omega_{\rm{sens}}(f)$ is the noise energy density power spectrum for the chosen detector. A signal is detectable if the observed SNR value exceeds a threshold SNR, denoted as $\rm{SNR_{thres}}$. The value of $\rm{SNR_{thres}}$ varies from detector to detector. We take $\rm{SNR_{thres}} = 1$ for the purpose of discussion.

Table\,\ref{table: bp} presents six benchmark points\,(BP) with high SNR values for FP-DECIGO, BBO, and Ultimate-DECIGO, obtained using eq.\,\eqref{eq: SNR}. BP1, BP2, BP3, and BP4 have been chosen at the $SU(2)_R$ breaking scale $v_R=30$ TeV, while for BP5 and BP6 the chosen scales are $v_R=20$ TeV and $50$ TeV respectively. The top segment of the table shows the values of the quartic couplings, while the middle segment gives the mass spectrum corresponding to each BP. The bottom segment gives the values of PT parameters $\a$, $\beta/H_*$, $T_c$ and $T_n$. Barring BP1, all other BPs have $w\sim\mathcal{O}(1)$. All BPs have $\rho_1\lesssim \mathcal{O}(10^{-1})$ and hence smaller values of $m_{H_3}$ are preferred. 

\begin{figure}[tbp]
\centering 
\includegraphics[width=.8\textwidth]{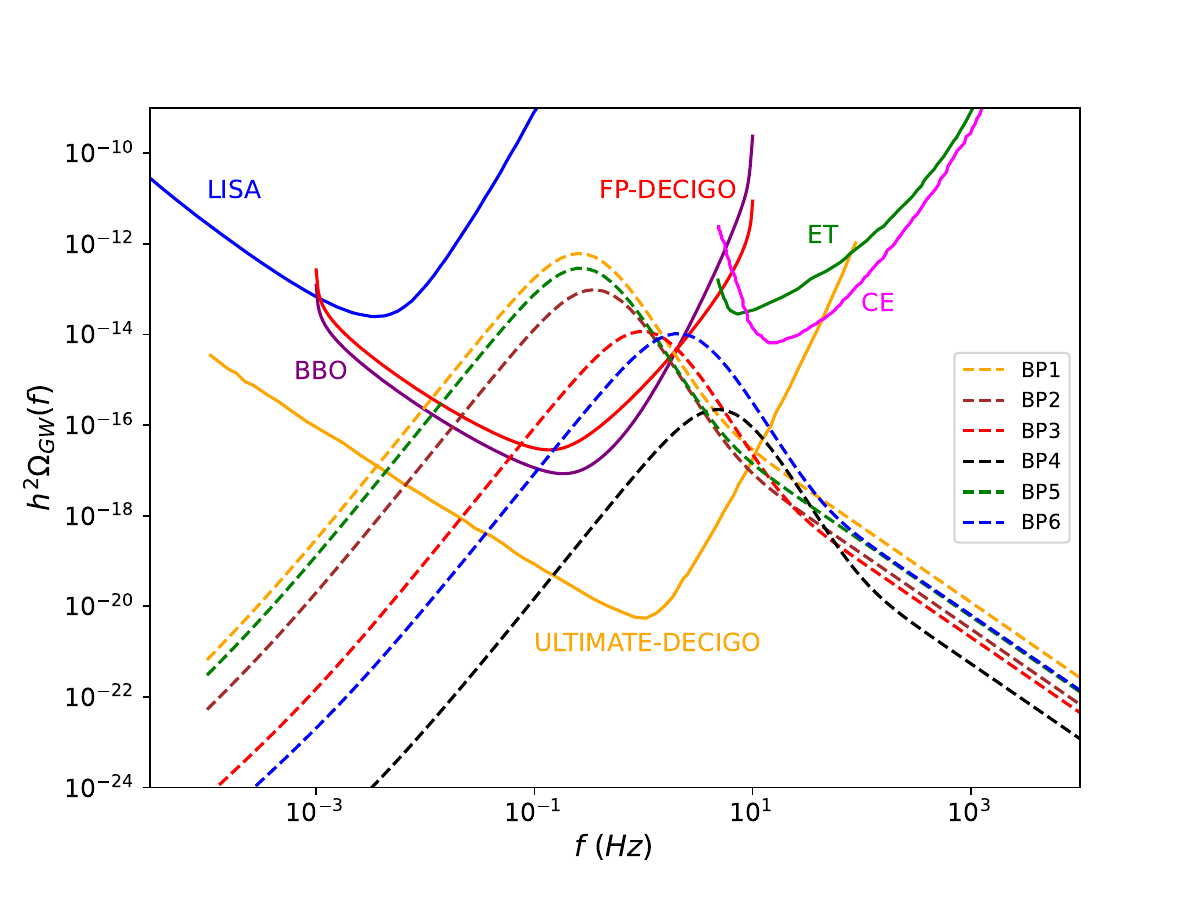}
\caption{\label{fig: benchmarks} GW spectra for the benchmark points listed in table\,\ref{table: bp}.} 
\end{figure}

The full GW spectra for the BPs are shown in fig.\,\ref{fig: benchmarks}. The peak of the spectrum corresponds to the frequency $f_{\rm{sw}}$ defined in eq.\,\eqref{eq: f_sw} since $\Omega_{\rm{sw}}$ gives the dominant contribution. The peak of BP4 lies only above Ultimate-DECIGO and below BBO and FP-DECIGO, while all other BPs have GW peaks above the sensitivity curves of Ultimate-DECIGO, BBO, and FP-DECIGO. The low- and high-frequency tails are dominated by the power law behavior of $\Omega_{\rm{turb}}$.

The SNR of the BPs are listed in table\,\ref{table: SNR}. As proclaimed in the previous section, the BPs generally yield high SNR values for FP-DECIGO, BBO, and Ultimate-DECIGO. The SNR values for BP1, BP2, BP3, BP5 and BP6 are higher than $1$ for FP-DECIGO, BBO, and Ultimate-DECIGO, and hence have good detection prospects. Ultimate-DECIGO, being the most sensitive, can detect all the BPs listed in table\,\ref{table: SNR} with large SNR values $>10^4$. The point BP4 is not detectable at FP-DECIGO and BBO, but can be detected by Ultimate-DECIGO.

\vfill

\begin{table}[tbp] 
\begin{center}
\vspace{0.5 cm}
\begin{tabular}{|c|c c c c c c|}
\hline
 SNR     & BP1             & BP2              & BP3                & BP4                 & BP5              & BP6                \\ 
\hline
FP-DECIGO   & $6.5\times10^3$ & $736.0$  & $14.4$    &   $6.5\times 10^{-3}$     & $3.0\times10^3$  & $ 2.4$      \\  
BBO      & $5.4\times 10^4$ &  $7.0\times 10^3$  &  $174.2$  & $6.5\times 10^{-2}$    &  $2.5\times 10^4$ &  $28.2$  \\ 
Ultimate-DECIGO  & $1.2\times 10^9$ & $2.6\times 10^8$ & $2.9\times 10^7$ & $2.2\times 10^4$   & $6.0\times 10^8$  & $8.0\times 10^6$ \\
\hline
\end{tabular}
\end{center}
\caption{\label{table: SNR} SNR values corresponding to different detectors for the benchmark points.
}
\end{table}

\section{Complementary collider probes}
\label{Sec: collider}
Now we describe the collider probes that complement the GW signatures discussed in the previous sections. We discuss two important collider implications, namely the  precision of $\k_h$ and detection of $H_3$.

\begin{table}[tbp] 
\begin{center}
\vspace{0.5 cm}
\begin{tabular}{|c|c|c|c|c|}
\hline
$ \delta \k_{h}$ & $20$ TeV & $30$ TeV & $50$ TeV  & Combined      \\ 
\hline
$> 5\%$ & $52\%$  & $58\%$ & $50\%$ & $ 54\%$  \\  
$>10\%$ & $21\%$  & $34\%$ & $33\%$ & $ 30\%$  \\  
$>20\%$ & $8\%$   & $20\%$ & $25\%$ & $ 17\%$  \\ 
$>50\%$ & $1.3\%$ & $12\%$ & $15\%$ & $ 9\%$   \\
\hline
\end{tabular}
\end{center}
\caption{\label{table: khhh} 
Percentage of points detectable at Ultimate-DECIGO to be ruled out when the sensitivity of $\k_h$ reaches $5\%, 10\%, 20\%,$ and $50\%$, for $v_R = 20, 30,$ and $50\,$TeV. 
}
\end{table}

\begin{itemize}
\item As argued in Sec.\,\ref{subsec: higgs}, in DLRSM the trilinear Higgs coupling can deviate significantly from its SM value. In Table\,\ref{table: khhh}, we present the percentage of points leading to detectable GW signal at Ultimate-DECIGO, which also show deviation of $\k_h$ at $5\%,~10\%,~20\%,$ and $50\%$. The current ATLAS measurement allows for a rather large range of $\k_h\in[-2.3,10.3]$. However, future colliders will significantly tighten the bound. Here we quote the projected sensitivities of $\k_h$ from ref.\,\cite{deBlas:2019rxi}. HL-LHC will achieve a sensitivity of $50\%$ from the di-Higgs production channel. The proposed colliders, such as HE-LHC, CLIC$_{3000}$, and FCC-hh are expected to improve the sensitivity of $\k_h$ to $\sim 20\%, 10\%,$ and $5\%$ respectively. These colliders therefore will rule out a considerable number of points showing a strong GW signal.

\begin{figure}[tbp]
\centering 
\includegraphics[width=.8\textwidth]{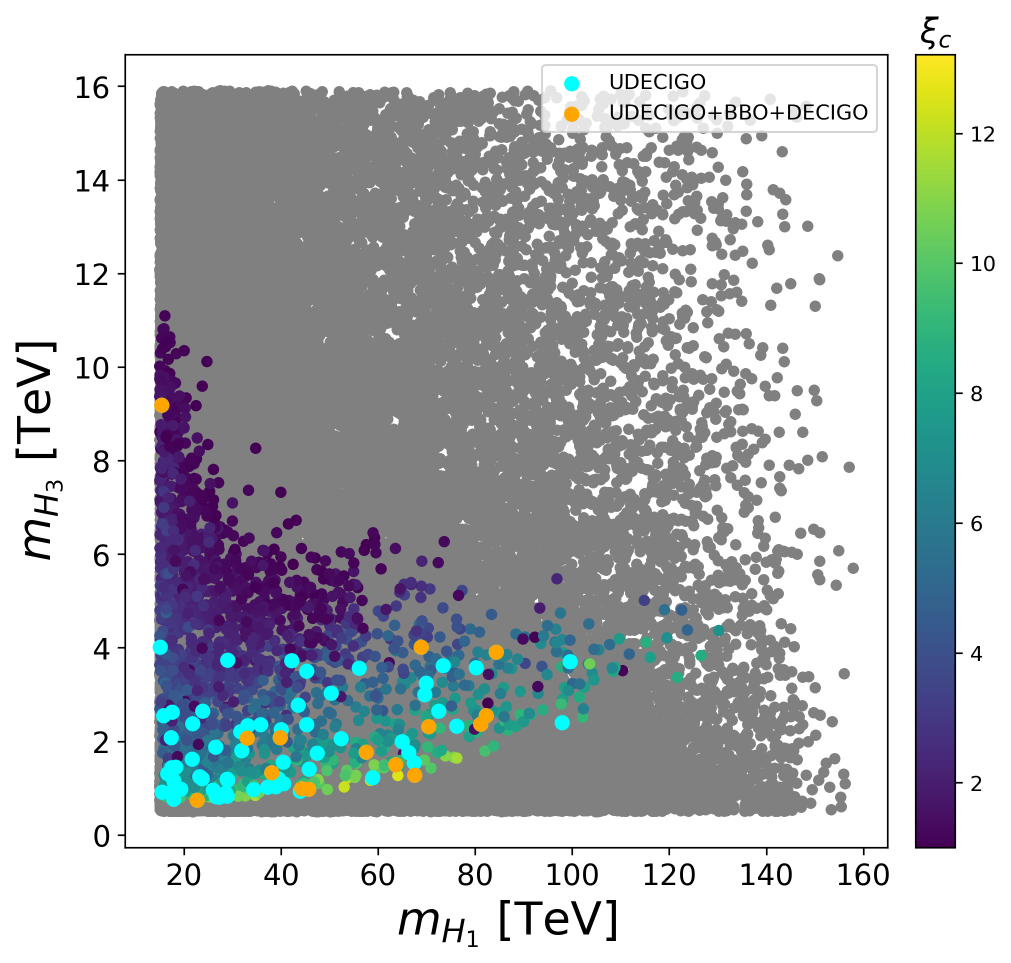}
\caption{\label{fig: mH1_mH3} The mass spectrum of DLRSM for $v_R = 20$\,TeV depicting points with $\xi_c > 1$. The cyan and orange points lead to a GW signal detectable at Ultimate-DECIGO and Ultimate-DECIGO+BBO+FP-DECIGO respectively.}
\label{fig:rwplane}
\end{figure}

\item The scalar $H_3$ can be produced at $pp$ colliders through several channels, for example\,\cite{Dev:2016dja}, 
\begin{enumerate}[(i)]
    \item $H_1$-decay, $p p \rightarrow H_1 \rightarrow h H_3$,
    \item decay of boosted $h$, $p p \rightarrow h^* \rightarrow h H_3, H_3 H_3$,
    \item Higgsstrahlung, $p p \rightarrow V_R^{*} \rightarrow V_R H_3$,
    \item $V_R V_R$ fusion, $p p \rightarrow H_3 jj$\,\,.
\end{enumerate}
The relative strength of these processes depends on the mass spectrum of DLRSM. In fig.\,\ref{fig: mH1_mH3}, we show the distribution of SFOPT points in the $m_{H_1} - m_{H_3}$ plane for $v_R = 20$\,TeV, overlaid with points which are detectable at Ultimate-DECIGO, BBO, and FP-DECIGO. The detectable points mostly occur for small $m_{H_3}$, with the minimum value of $m_{H_3} = 741 $\,GeV. For the range $m_{H_3} = 740$\,GeV $-$ 1.2\,TeV, the production cross-section of $H_3$ at FCC-hh with $\sqrt{s} = 100$\,TeV can be $\sim \mathcal{O}(\text{fb})$\,\cite{Dev:2016dja}.

For $v_R = 20$\,TeV, $m_{H_3} \lesssim 500$\,GeV can be ruled out from the observations of the channel\,(iv) at FCC-hh with a luminosity of $30\,\text{ab}^{-1}$. For large values of quartic couplings, the decay width $h^{*} \rightarrow h H_3$ and $h^{*} \rightarrow H_3 H_3$ can be large and subsequently, channel (ii) can rule out $m_{H_3} \lesssim 700$\,GeV. For channel (i), $H_1$ with mass $15$\,TeV can be produced with a cross-section $\sim 0.5$\,fb and have sizable branching ratios of $H_1 \rightarrow h H_3,\, H_3 H_3$. As a result, channel (i) can rule out masses up to $m_{H_3} \sim 2$\,TeV. Thus, these searches are capable of ruling out a large number of points with low-$m_{H_3}$, thus low-$\rho_1$, providing a complementarity to the GW probe of DLRSM. 

\end{itemize}
\section{Summary and conclusions}\label{sec: summary}
In this paper, we studied the possibility of an observable stochastic GW background resulting from SFOPT associated with the spontaneous breaking of $SU(2)_R\times U(1)_{B-L}$ in DLRSM. The gauge symmetry of DLRSM breaks in the following pattern:
$$SU(2)_L\times SU(2)_R \times U(1)_{B-L}\xrightarrow{\mathit{~~v_R~~}} SU(2)_L\times U(1)_{Y} \xrightarrow{\mathit{\k_1,\k_2,v_L}} U(1)_Y . $$
The non-observation of a right-handed current at colliders puts a lower bound on the scale $v_R$ to be around 20 TeV. Due to the hierarchy $v_R \gg v$, the $SU(2)_R\times U(1)_{B-L}$-breaking dynamics is decoupled from the EWPT.  We chose the scale $v_R=20,~30,$ and $50$ TeV to study the possible detection of GW background at the planned observatories. For these values of $v_R$, complementary searches for new scalars of DLRSM are feasible at future colliders. 

Our analysis was carried out using the {\textit{simple basis}} defined in ref.\,\cite{Karmakar:2022iip}, to reduce the number of independent parameters. It should be noted that analysis with the full set of parameters also gives similar patterns of SFOPT in the $\rho_1-\rho_2$ and $r-w$ planes. The parameters in the {\textit{simple basis}} include the quartic couplings: $\l_0, \l_2, \a_0, \a_3, \rho_1, \rho_2$. In addition, we defined EW \vevs through the ratios $r$ and $w$. Most studies on LRSM take the simplified limit $r,w\rightarrow 0$. However, it was pointed out in refs.\,\cite{Bernard:2020cyi,Karmakar:2022iip} that the DLRSM phenomenology allows for significant deviation from this limit. Therefore, we also scanned over $r$ and $w$. 

We constructed the one-loop finite temperature effective potential for each parameter point and analyzed the nature of PT using the package \texttt{CosmoTransitions}. Due to the large separation between $v_R$ and the EW scale, the effective potential depends solely on the background field value of the neutral CP-even scalar, $\chi^0_{Rr}$. The condition for SFOPT, $\xi_c>1$ was used to identify viable regions of the parameter space. SFOPT favors small values of the quartic coupling $\rho_1\lesssim \mathcal{O}(10^{-1})$, which leads to $m_{H_3} \ll v_R$. This feature has also been observed in other variants of LRSM, discussed in refs.\,\cite{Brdar:2019fur, Graf:2021xku, Li:2020eun}. 

We find that for very small values, $\rho_1\lesssim 10^{-3}$ however, the zero temperature minimum of the one-loop effective potential at $R = v_R$ becomes metastable, violating the Linde-Weinberg bound. Hence there is a lower bound on $\rho_1$ below which FOPT is not observed. Most points with SFOPT also feature $w\sim \mathcal{O}(1)$, while for smaller values of $w$, very few points show SFOPT. Out of the chosen set of parameters, the SFOPT region is most sensitive to the parameters $\rho_1$ and $w$ and to some extent $\l_0$. However, we see no particular preference for the \vev ratio $r$ and the quartic couplings relating the bidoublet and the doublet fields, i.e., $\a_0$ and $\a_3$, as illustrated by the projections given in fig.\,\ref{fig: projections}. 

For parameter points showing SFOPT, we computed the PT parameters, $\a$, $\beta/H_*$, and $T_n$, needed for the calculation of the GW spectrum. In the non-runaway scenario, the stochastic GW background resulting from SFOPT comes primarily from sound waves and turbulence, while the contribution from bubble wall collisions remains sub-dominant. Fig.\,\ref{fig: gw_scatter} shows the position of the peak of the GW spectrum for points satisfying the SFOPT criterion. While for a large number of points, the GW spectrum is too weak to be detected, there is a significant number of points lying above the sensitivity curves for Ultimate-DECIGO, BBO, and FP-DECIGO. Such points will be accessible to these detectors in the coming years. The detectable points also prefer $w \sim \mathcal{O}(1)$, which in turn, correspond to a large value of $y_{33}$ as seen in fig.\,\ref{fig: rw_203050}. 

The strength of the GW spectrum does not depend on the scale $v_R$. On the other hand, since the peak frequency is proportional to $T_n\sim v_R$, the points shift to the right as $v_R$ changes from $20$ to $30$ to $50$ TeV. To quantify the detection prospects, we computed the signal-to-noise ratio at these detectors for the detectable points. Six benchmark points are given in table\,\ref{table: SNR}, featuring SNR values higher than $10^5$. We see that for all the BPs, $m_{H_3}\lesssim 5$ TeV. 

There are primarily two complementary collider probes for the points with detectable GW signals. It was found that a significant fraction of points leads to $50, 20, 10, \text{and}\, 5\%$ deviation of $k_h$ from unity, which can be ruled out at HL-LHC, HE-LHC, CLIC$_{3000}$, and FCC-hh respectively. Due to a relatively low mass of $H_3$, it can be produced at future colliders through various channels. In particular, FCC-hh can rule out up to $m_{H_3} \sim 2$\,TeV.  

Here we make a note of some subtelties involved in computing the GW spectrum that contribute to theoretical uncertainty: \textbf{(i)} The suppression factor $\Upsilon$, introduced in \eqref{eq: soundwaves} was recently proposed, to take the finite lifetime of sound waves into account. For the chosen benchmark points this suppression factor takes $\mathcal{O}(0.1)$ values. \textbf{(ii)} As pointed out earlier, the value of $v_c/T_c$ depends on the particular choice of gauge since the effective potential is gauge dependent. The effect of gauge-dependence in minimized in the Landau gauge, which we use for our calculations. \textbf{(iii)} In principle, the bubble wall velocity can be computed from the model parameters, as seen in refs.\,\cite{Moore:1995ua,Moore:1995si,Bodeker:2009qy}. We use $v_w=1$, which is valid when the friction on the walls is low. Thus a particular choice of $v_w$ can cause small shifts in the GW spectrum. We checked that the uncertainties mentioned above contribute to roughly $\mathcal{O}(0.1-1)$ deviations in the GW spectrum of the BPs. However, the BPs would still be detectable at respective detetectors, BBO and/or DECIGO.

The spontaneous breaking of the discrete LR symmetry $\mathcal{P}$ can lead to the formation of domain walls. The GW imprint from the domain wall network peaks at much lower frequencies, as compared to that from FOPT \cite{Borah:2022wdy, Borboruah:2022eex}. Since there is no  overlap of the GW signals, we have focussed our discussion on FOPT. 

Although DLRSM does not account for neutrino masses, it is interesting to ask if incorporating them by adding extra fields to the model could modify the strength of FOPT. In Appendix\,\ref{sec: neutrino mass}, we have shown that it is possible to include neutrino masses without impacting the results of our analysis.

\section*{Acknowledgments}
DR is thankful to Subhendu Rakshit for his useful suggestions. SK acknowledges discussions with S. Uma Sankar during an earlier collaboration. This research work uses the computing facilities under DST-FIST scheme\,(Grant No. SR/FST/PSI-225/2016) of the Department of Science and Technology\,(DST), Government of India.
DR is thankful for the support from 
DST, via SERB
Grants no.\,MTR/2019/000997 and no.\,CRG/2019/002354. DR is supported by the Government of India
UGC-SRF fellowship. 
SK thanks  DTP, TIFR Mumbai for funding the Visiting Fellow position, where part of the work was completed. 

\appendix
\section{Minimization at the EW vacua}\label{appendix: min}
The minimization conditions are given by:
\bea \label{eq: minimization}
\mu_1^2 &=& \frac{1}{2(r^2 -1)}\Bigg(\k_1^2 \Big(w^2((r^2-1)\a_1 + r^2 \a_3 - \a_4) + 2 (r^2 -1)((r^2 + 1)\l_1 + 2 r \l_4) \Big) \nn\\
&& + 2 \sqrt{2} r v_R w \m_4 + v_R^2 \Big((r^2 -1)\a_1 + r^2 \a_3 - \a_4  + 2 w^2 \rho_{12}\Big) \Bigg) \,\,, \nn\\
\mu_2^2 &=&  \frac{1}{4(r^2 -1)}\Bigg( \k_1^2 \Big(w^2 (r^2-1) \a_2 - w^2 r \a_{34} + 2 (r^2 -1)(2 r \l_{23} + (r^2+1)\l_4) \Big)\nn\\
&& - \sqrt{2}(r^2+1) v_R w \m_4 + v_R^2 \Big((r^2-1)\a_2 - r \a_{34} - 2 w^2 \rho_{12} \Big) \Bigg)\,\,, \nn\\
\m_3^2 & = & \frac{1}{2} \k_1^2 ((r^2+1)\a_1 + 2 r \a_2 + r^2 \a_3 + \a_4 + 2 w^2 \rho_{1}) + v_R^2 \rho_1\,\,, \nn\\
\m_5 &=& - r \m_4 - \sqrt{2} v_R w \rho_{12},
\label{eq:minimisationcondition}
\eea
 where, $\rho_{12}=\rho_2/2 -\rho_1$, $\a_{34}=\a_3-\a_4$, and $\l_{23} = 2 \l_2 + \l_3$.  
\section{Field-dependent masses}\label{appendix: field}
The field-dependent mass matrices are obtained from the tree-level effective potential:
\beq
    m^2_{ij}(R) = \left.\frac{\partial^2}{\partial \varphi_i \partial \varphi_j} V_0\right\vert_{\langle \cdots \rangle}
\eeq
where $\langle \cdots \rangle$ denotes the background field value. This amounts to replacing $v_R\rightarrow R$, and $\k_1,\k_2,v_L\rightarrow 0$ in the usual mass matrices.

For the CP-even sector, in the basis $\{\phi_{1r}^0,\phi_{2r}^0,\chi_{Lr}^0,\chi_{Rr}^0\}$, we obtain,
\beq
    \mathcal{M}^2_{\rm{CPE}} = \begin{pmatrix}
        -\mu_1^2 + \frac{1}{2}(\alpha_1+\alpha_4)R^2 & -2\mu_2^2+\frac{1}{2}\alpha_2 R^2 & \frac{1}{\sqrt{2}}\mu_5 R & 0\\
        -2\mu_2^2+\frac{1}{2}\alpha_2R^2 & -\mu_1^2 + \frac{1}{2}(\alpha_1+\alpha_3)R^2 & \frac{1}{\sqrt{2}}\mu_4 R & 0\\
        \frac{1}{\sqrt{2}}\mu_5 R & \frac{1}{\sqrt{2}}\mu_4 R & -\mu_3^2+\frac{1}{2} \rho_2 R^2 & 0 \\
        0 & 0 & 0 & -\mu_3^2+ 3\rho_1 R^2\\
    \end{pmatrix}.
\eeq
For the CP-odd scalars, in the basis $\{\phi_{1i}^0,\phi_{2i}^0,\chi_{Li}^0,\chi_{Ri}^0\}$,
\beq
    \mathcal{M}^2_{\rm{CP0}} = \begin{pmatrix}
        -\mu_1^2 + \frac{1}{2}(\alpha_1+\alpha_4)R^2 & 2\mu_2^2-\frac{1}{2}\alpha_2R^2 & -\frac{1}{\sqrt{2}}\mu_5 R & 0\\
        2\mu_2^2-\frac{1}{2}\alpha_2R^2 & -\mu_1^2 + \frac{1}{2}(\alpha_1+\alpha_3)R^2 & \frac{1}{\sqrt{2}}\mu_4 R & 0\\
        -\frac{1}{\sqrt{2}}\mu_5 R & \frac{1}{\sqrt{2}}\mu_4 R & -\mu_3^2+\frac{1}{2} \rho_2 R^2 & 0 \\
        0 & 0 & 0 & -\mu_3^2+ \rho_1 R^2\\
    \end{pmatrix},
\eeq
and for the charged scalars, in the basis $\{\phi_{1}^{\pm},\phi_{2}^{\pm},\chi_{L}^{\pm},\chi_{R}^{\pm}\}$ we get,
\beq
    \mathcal{M}^2_{\rm{charged}} = \begin{pmatrix}
        -\mu_1^2 + \frac{1}{2}(\alpha_1+\alpha_4)R^2 & 2\mu_2^2-\frac{1}{2}\alpha_2R^2 & -\frac{1}{\sqrt{2}}\mu_5 R & 0\\
        2\mu_2^2-\frac{1}{2}\alpha_2R^2 & -\mu_1^2 + \frac{1}{2}(\alpha_1+\alpha_3)R^2 & \frac{1}{\sqrt{2}}\mu_4 R & 0\\
        -\frac{1}{\sqrt{2}}\mu_5 R & \frac{1}{\sqrt{2}}\mu_4 R & -\mu_3^2+\frac{1}{2} \rho_2 R^2 & 0 \\
        0 & 0 & 0 & -\mu_3^2+ \rho_1 R^2\\
    \end{pmatrix}.
\eeq

The neutral gauge boson mass matrix, in the basis $\{Z_L^{\mu},Z_R^{\mu},B^{\mu}\}$, is,
\beq
    \mathcal{M}^2_{Z} = \begin{pmatrix}
        0 & 0 & 0\\
        0 & 0 & 0\\
        0 & \frac{1}{4}g_R^2 R^2 & -\frac{1}{4}g_{BL}g_R R^2\\
        0 & -\frac{1}{4}g_{BL}g_R R^2 & \frac{1}{4}g_{BL}^2 R^2\\
    \end{pmatrix}.
\eeq

For the charged bosons, in the basis $\{W_L^{\mu\pm},W_R^{\mu\pm}\}$,
\beq
    \mathcal{M}^2_{W} = \begin{pmatrix}
        0 & 0\\
        0 & \frac{1}{4}g_R^2 R^2\\
    \end{pmatrix}.
\eeq

In addition to the field-dependent masses, we also need thermal self-energies of the fields for daisy resummation. These are obtained from the high-T expansion of the one-loop thermal potential. Substituting eq.\,\eqref{eq: highTb}  in eq.\,\eqref{eq: V1T} gives, to leading order,
\beq
    V_{1T}^{\rm{high}} = \frac{T^2}{24}\left(\sum_b n_b m_b^2 + \frac{1}{2}\sum_f n_f m_f^2\right).
\eeq
Here, index $b$ runs over bosons, while index $f$ runs over fermions. Each sum can be expressed as the trace of the respective matrix. Thermal mass matrices are then expressed as, $\Pi_{ij} = c_{ij} T^2$, where $c_{ij}$ are,
\beq
    c_{ij} = \frac{1}{T^2}\left.\frac{\partial^2}{\partial\varphi_i\partial\varphi_j}V_{1T}^{\rm{high}}\right\vert_{\langle\cdots\rangle}.
\eeq
We define,
\bea
     d_1 &=&\frac{1}{48}(9 g_L^2 + 9 g_R^2 + 8(2\alpha_1+\alpha_3+\alpha_4+5\lambda_1+2\lambda_3)),\\
     d'_1 &=& d_1 + \frac{y_{33}^2}{4} + \frac{\tilde{y}_{33}^2}{4},\\
     d_2 &=& \frac{1}{3}(2\alpha_2+3\lambda_4),\\
     d'_2 &=& d_2 + \frac{y_{33}\tilde{y}_{33}}{4},\\
     d_L &=& \frac{1}{48}(3 g_{BL}^2 + 9 g_L^2 + 8(2\alpha_1+\alpha_3+\alpha_4+3\rho_1+\rho_2)),\\
     d_R &=& \frac{1}{48}(3 g_{BL}^2 + 9 g_R^2 + 8(2\alpha_1+\alpha_3+\alpha_4+3\rho_1+\rho_2))\,.
\eea
We obtain the following thermal mass matrices:
\beq
    \Pi_{\rm{CPE}} = T^2 \begin{pmatrix}
        d'_1 & d'_2 & 0 & 0\\
        d'_2 & d'_1 & 0 & 0\\
        0  & 0 & d_L & 0 \\
        0 & 0 & 0 & d_R\\
    \end{pmatrix},
\eeq
\beq
    \Pi_{\rm{CP0}} = T^2 \begin{pmatrix}
        d'_1 & -d'_2 & 0   & 0\\
      -d'_2 & d'_1   & 0   & 0\\
         0  & 0     & d_L & 0 \\
         0  & 0     & 0   & d_R\\
    \end{pmatrix},
\eeq
\beq
    \Pi_{\rm{charged}} = T^2 \begin{pmatrix}
        d_1 & -d_2 & 0 & 0\\
        -d_2 & d_1 & 0 & 0\\
        0  & 0 & d_L & 0 \\
        0 & 0 & 0 & d_R\\
    \end{pmatrix}.
\eeq

The thermal mass matrices for the longitudinal gauge bosons are, 
\beq
    \Pi_Z = \frac{T^2}{6}\begin{pmatrix}
        13 g_L^2 & 0 & 0 \\
        0 & 13 g_R^2 & 0 \\
        0 & 0 & 6 g_{BL}^2\\
    \end{pmatrix}, 
\eeq
\beq
    \Pi_{W^{\pm}} = \frac{13}{6}T^2\begin{pmatrix}
        g_L^2 & 0 \\
        0 &  g_R^2 \\
    \end{pmatrix}.
\eeq
The mass of each species with the above thermal corrections is obtained as the eigenvalue of the matrix, $m^2(R)+\Pi(T)$. After diagonalization, the longitudinal polarization of the photon becomes massive, while the transverse components remain massless. 

\section{Neutrino masses in DLRSM}\label{sec: neutrino mass}
We have not taken into account a mechanism for generating neutrino mass in our version of DLRSM.  In this section, we argue the minimal way of incorporating neutrino mass in this model do not give any additional contribution to the GW phenomenology of the model. 
%rewrite if necessary

To demonstrate our point, we consider the model discussed in refs.\,\cite{FileviezPerez:2016erl,FileviezPerez:2017zwm}. Small neutrino masses are generated radiatively by the Zee mechanism, by adding a charged singlet scalar $\delta^+\sim(1,1,1,2)$ to DLRSM. In our notation, the Majorana Lagrangian is,
\beq
-\mathcal{L}^M_{LR} = \gamma_L L_L L_L\delta^+ + \gamma_R L_R L_R\delta^+ + \gamma_1 \chi_L^T i\sigma_2\Phi\chi_R\delta^- + \gamma_1 \chi_L^T i\sigma_2\tilde{\Phi}\chi_R\delta^- + \text{h.c.}\,,
\eeq
where, $\gamma_{L,R},~\gamma_1,~\gamma_2$ are the new Yukawa couplings.
As there is no tree-level right-handed neutrino mass, the contribution of the RH neutrinos to the effective potential is zero.  However, the quartic terms involving $\delta^+$ modify the mixing between the charged scalars. In the basis of $\{\phi_1^{\pm},\phi_2^{\pm},\chi_L^{\pm},\chi_R^{\pm},\delta^{\pm}\}$, the additional contribution to the charged mass matrix, $\mathcal{M}^2_{\rm_{charged}}$ is, 
\beq
\delta \mathcal{M}^2_{\rm_{charged}} = v_R^2\begin{pmatrix}
0 & 0 & 0 & 0 & \frac{\gamma_2}{2} \frac{v_L}{v_R}\\
0 & 0 & 0 & 0 & -\frac{\gamma_1}{2} \frac{v_L}{v_R}\\
0 & 0 & 0 & 0 & \frac{(\gamma_1\k_2+\gamma_2\k_1)}{2 v_R} \\
0 & 0 & 0 & 0 & -\frac{v_L(\gamma_1\k_1+\gamma_2\k_2)}{2v_R^2}\\
\frac{\gamma_2}{2} \frac{v_L}{v_R} & -\frac{\gamma_1}{2} \frac{v_L}{v_R} & \frac{(\gamma_1\k_2+\gamma_2\k_1)}{2 v_R}  & -\frac{v_L(\gamma_1\k_1+\gamma_2\k_2)}{2v_R^2}  & 0
\end{pmatrix}.
\eeq

Each of the non-zero entries is suppressed by a factor $v_L/v_R$ or $\k_{1,2}/v_R$ compared to $v_R^2$. Therefore the mixing of the charged scalars of DLRSM with $\delta^+$ is negligible, while their mixing among themselves remains unchanged. In the field-dependent mass matrix, we put $v_L\rightarrow 0,~\k_{1,2}\rightarrow0$, and $v_R\rightarrow R$, by which the additional mixing matrix, $\delta \mathcal{M}^2_{\rm_{charged}}(R)$, vanishes entirely. Hence the presence of $\delta^+$ does not alter the field-dependent mass matrices and therefore does not contribute to the effective potential. 

\bibliographystyle{unsrt}
\bibliography{citation.bib}
\end{document}